\documentclass[aps,prc,superscriptaddress,showpacs,floatfix,letterpaper,preprintnumbers,nofootinbib,twocolumn]{revtex4-2}
\usepackage{epsfig}
\usepackage{bm}
\usepackage{amsmath}
\usepackage{amstext}
\usepackage{amssymb}
\usepackage{mathrsfs}
\usepackage{color}
\usepackage{subfigure}
\usepackage{graphicx}
\usepackage{hyperref}
\hypersetup{
    colorlinks=true,
    linkcolor=blue,
    filecolor=magenta,      
    urlcolor=blue,
    citecolor=blue
}

\newcommand{\trento}{\texttt{T$_R$ENTo} }

\begin{document}

\title{A new metric improving Bayesian calibration of a multistage approach studying hadron and inclusive jet suppression}

\author{W.~Fan}
\email[Corresponding author: ]{wenkai.fan@duke.edu}
\affiliation{Department of Physics, Duke University, Durham NC 27708.}

\author{G.~Vujanovic}
\email[Corresponding author: ]{gojko.vujanovic@uregina.ca}
\affiliation{Department of Physics and Astronomy, Wayne State University, Detroit MI 48201.}
\affiliation{Department of Physics, University of Regina, Regina, SK S4S 0A2, Canada}

\author{S.~A.~Bass}
\affiliation{Department of Physics, Duke University, Durham NC 27708.}




\author{A.~Angerami}
\affiliation{Lawrence Livermore National Laboratory, Livermore CA 94550.}

\author{R.~Arora}
\affiliation{Research Computing Group, University Technology Solutions, The University of Texas at San Antonio, San Antonio TX 78249.}


\author{S.~Cao}
\affiliation{Institute of Frontier and Interdisciplinary Science, Shandong University, Qingdao, Shandong 266237, China}
\affiliation{Department of Physics and Astronomy, Wayne State University, Detroit MI 48201.}

\author{Y.~Chen}
\affiliation{Laboratory for Nuclear Science, Massachusetts Institute of Technology, Cambridge MA 02139.}
\affiliation{Department of Physics, Massachusetts Institute of Technology, Cambridge MA 02139.}

\author{T.~Dai}
\affiliation{Department of Physics, Duke University, Durham NC 27708.}

\author{L.~Du}
\affiliation{Department of Physics, McGill University, Montr\'{e}al QC H3A\,2T8, Canada.}

\author{R.~Ehlers}
\affiliation{Department of Physics and Astronomy, University of Tennessee, Knoxville TN 37996.}
\affiliation{Physics Division, Oak Ridge National Laboratory, Oak Ridge TN 37830.}

\author{H.~Elfner}
\affiliation{GSI Helmholtzzentrum f\"{u}r Schwerionenforschung, 64291 Darmstadt, Germany.}
\affiliation{Institute for Theoretical Physics, Goethe University, 60438 Frankfurt am Main, Germany.}
\affiliation{Frankfurt Institute for Advanced Studies, 60438 Frankfurt am Main, Germany.}

\author{R.~J.~Fries}
\affiliation{Cyclotron Institute, Texas A\&M University, College Station TX 77843.}
\affiliation{Department of Physics and Astronomy, Texas A\&M University, College Station TX 77843.}

\author{C.~Gale}
\affiliation{Department of Physics, McGill University, Montr\'{e}al QC H3A\,2T8, Canada.}


\author{Y.~He}
\affiliation{Guangdong Provincial Key Laboratory of Nuclear Science, Institute of Quantum Matter, South China Normal University, Guangzhou 510006, China.}
\affiliation{Guangdong-Hong Kong Joint Laboratory of Quantum Matter, Southern Nuclear Science Computing Center, South China Normal University, Guangzhou 510006, China.}

\author{M.~Heffernan}
\affiliation{Department of Physics, McGill University, Montr\'{e}al QC H3A\,2T8, Canada.}

\author{U.~Heinz}
\affiliation{Department of Physics, The Ohio State University, Columbus OH 43210.}

\author{B.~V.~Jacak}
\affiliation{Department of Physics, University of California, Berkeley CA 94270.}
\affiliation{Nuclear Science Division, Lawrence Berkeley National Laboratory, Berkeley CA 94270.}

\author{P.~M.~Jacobs}
\affiliation{Department of Physics, University of California, Berkeley CA 94270.}
\affiliation{Nuclear Science Division, Lawrence Berkeley National Laboratory, Berkeley CA 94270.}

\author{S.~Jeon}
\affiliation{Department of Physics, McGill University, Montr\'{e}al QC H3A\,2T8, Canada.}

\author{Y.~Ji}
\affiliation{Department of Statistical Science, Duke University, Durham NC 27708.}


\author{L.~Kasper}
\affiliation{Department of Physics and Astronomy, Vanderbilt University, Nashville TN 37235.}




\author{M.~Kordell~II}
\affiliation{Cyclotron Institute, Texas A\&M University, College Station TX 77843.}
\affiliation{Department of Physics and Astronomy, Texas A\&M University, College Station TX 77843.}

\author{A.~Kumar}
\affiliation{Department of Physics, McGill University, Montr\'{e}al QC H3A\,2T8, Canada.}
\affiliation{Department of Physics and Astronomy, Wayne State University, Detroit MI 48201.}

\author{J.~Latessa}
\affiliation{Department of Computer Science, Wayne State University, Detroit MI 48202.}

\author{Y.-J.~Lee}
\affiliation{Laboratory for Nuclear Science, Massachusetts Institute of Technology, Cambridge MA 02139.}
\affiliation{Department of Physics, Massachusetts Institute of Technology, Cambridge MA 02139.}

\author{R.~Lemmon}
\affiliation{Daresbury Laboratory, Daresbury, Warrington, Cheshire, WA44AD, United Kingdom.}

\author{D.~Liyanage}
\affiliation{Department of Physics, The Ohio State University, Columbus OH 43210.}

\author{A.~Lopez}
\affiliation{Instituto  de  F\`{i}sica,  Universidade  de  S\~{a}o  Paulo,  C.P.  66318,  05315-970  S\~{a}o  Paulo,  SP,  Brazil. }

\author{M.~Luzum}
\affiliation{Instituto  de  F\`{i}sica,  Universidade  de  S\~{a}o  Paulo,  C.P.  66318,  05315-970  S\~{a}o  Paulo,  SP,  Brazil. }

\author{A.~Majumder}
\affiliation{Department of Physics and Astronomy, Wayne State University, Detroit MI 48201.}

\author{S.~Mak}
\affiliation{Department of Statistical Science, Duke University, Durham NC 27708.}

\author{A.~Mankolli}
\affiliation{Department of Physics and Astronomy, Vanderbilt University, Nashville TN 37235.}

\author{C.~Martin}
\affiliation{Department of Physics and Astronomy, University of Tennessee, Knoxville TN 37996.}

\author{H.~Mehryar}
\affiliation{Department of Computer Science, Wayne State University, Detroit MI 48202.}

\author{T.~Mengel}
\affiliation{Department of Physics and Astronomy, University of Tennessee, Knoxville TN 37996.}

\author{J.~Mulligan}
\affiliation{Department of Physics, University of California, Berkeley CA 94270.}
\affiliation{Nuclear Science Division, Lawrence Berkeley National Laboratory, Berkeley CA 94270.}

\author{C.~Nattrass}
\affiliation{Department of Physics and Astronomy, University of Tennessee, Knoxville TN 37996.}

\author{J.~Norman}
\affiliation{Oliver Lodge Laboratory, University of Liverpool, Liverpool, United Kingdom.}


\author{J.-F.~Paquet}
\affiliation{Department of Physics, Duke University, Durham NC 27708.}

\author{C.~Parker}
\affiliation{Cyclotron Institute, Texas A\&M University, College Station TX 77843.}
\affiliation{Department of Physics and Astronomy, Texas A\&M University, College Station TX 77843.}

\author{J.~H.~Putschke}
\affiliation{Department of Physics and Astronomy, Wayne State University, Detroit MI 48201.}

\author{G.~Roland}
\affiliation{Laboratory for Nuclear Science, Massachusetts Institute of Technology, Cambridge MA 02139.}
\affiliation{Department of Physics, Massachusetts Institute of Technology, Cambridge MA 02139.}

\author{B.~Schenke}
\affiliation{Physics Department, Brookhaven National Laboratory, Upton NY 11973.}

\author{L.~Schwiebert}
\affiliation{Department of Computer Science, Wayne State University, Detroit MI 48202.}

\author{A.~Sengupta}
\affiliation{Cyclotron Institute, Texas A\&M University, College Station TX 77843.}
\affiliation{Department of Physics and Astronomy, Texas A\&M University, College Station TX 77843.}

\author{C.~Shen}
\affiliation{Department of Physics and Astronomy, Wayne State University, Detroit MI 48201.}
\affiliation{RIKEN BNL Research Center, Brookhaven National Laboratory, Upton NY 11973.}


\author{C.~Sirimanna}
\affiliation{Department of Physics and Astronomy, Wayne State University, Detroit MI 48201.}

\author{D.~Soeder}
\affiliation{Department of Physics, Duke University, Durham NC 27708.}

\author{R.~A.~Soltz}
\affiliation{Department of Physics and Astronomy, Wayne State University, Detroit MI 48201.}
\affiliation{Lawrence Livermore National Laboratory, Livermore CA 94550.}

\author{I.~Soudi}
\affiliation{Department of Physics and Astronomy, Wayne State University, Detroit MI 48201.}


\author{M.~Strickland}
\affiliation{Department of Physics, Kent State University, Kent, OH 44242.}

\author{Y.~Tachibana}
\affiliation{Akita International University, Yuwa, Akita-city 010-1292, Japan.}
\affiliation{Department of Physics and Astronomy, Wayne State University, Detroit MI 48201.}

\author{J.~Velkovska}
\affiliation{Department of Physics and Astronomy, Vanderbilt University, Nashville TN 37235.}

\author{X.-N.~Wang}
\affiliation{Key Laboratory of Quark and Lepton Physics (MOE) and Institute of Particle Physics, Central China Normal University, Wuhan 430079, China.}
\affiliation{Department of Physics, University of California, Berkeley CA 94270.}
\affiliation{Nuclear Science Division, Lawrence Berkeley National Laboratory, Berkeley CA 94270.}


\author{W.~Zhao}
\affiliation{Department of Physics and Astronomy, Wayne State University, Detroit MI 48201.}

\collaboration{The JETSCAPE Collaboration}

\begin{abstract}
We study parton energy-momentum exchange with the quark gluon plasma (QGP) within a multistage approach composed of in-medium DGLAP evolution at high virtuality, and (linearized) Boltzmann Transport formalism at lower virtuality. This multistage simulation is then calibrated in comparison with high $p_T$ charged hadrons, D-mesons, and the inclusive jet nuclear modification factors, using Bayesian model-to-data comparison, to extract the virtuality-dependent transverse momentum broadening transport coefficient $\hat{q}$. To facilitate this undertaking, we develop a quantitative metric for validating the Bayesian workflow, which is used to analyze the sensitivity of various model parameters to individual observables. The usefulness of this new metric in improving Bayesian model emulation is shown to be highly beneficial for future such analyses.      
\end{abstract}

\date{\today}
\maketitle
\section{Introduction}

Colliding QCD bound states at relativistic energies can lead to the excitation of its fundamental degrees of freedom known as partons. Some of these are highly energetic and generate a spray of particles known as a jet. While proton-proton collisions allow the study of the fragmentation of partons and their subsequent decay into hadronic bound states in the vacuum \cite{Field:1989uq, Skands:2014pea, ATLAS:2010osr,JETSCAPE:2019udz}, the showering of jets in high-energy heavy-ion collisions inherently includes interactions with the hot and dense nuclear medium known as the quark gluon plasma (QGP) \cite{Bjorken:1982tu,Appel:1985dq,Blaizot:1986ma,Wang:1992qdg,Gyulassy:1993hr,Baier:1996sk,Zakharov:1997uu,Wiedemann:2000za,Gyulassy:1999zd,Guo:2000nz,Jeon:2003gi, Arnold:2002ja, Djordjevic:2003zk,Djordjevic:2008iz,Majumder:2009ge}. The modification of jets in nucleus-nucleus ($A$-$A$) collisions compared to jets in proton-proton ($p$-$p$) collisions is referred to as jet quenching. As jets are generated early in heavy-ion collisions, their partonic content samples the properties of the QGP throughout its evolution and can be described by perturbative QCD, which provides us with an established approach to study them (see \cite{Majumder:2010qh, Cao:2020wlm, Connors:2017ptx,Majumder:2007iu} and references therein). Given that parton lifetime (or virtuality) plays an important role in how partons interact with the QGP, and no single Monte Carlo approach for parton energy loss has been devised that describes all virtualities at once, a multiscale approach is the preferred option, as explored herein.   

The interaction between an energetic parton and the QGP medium is divided into two regimes determined by the virtuality ($t=E^2-|{\bf p}|^2=Q^2$) of the parton. Parton evolution in the high-virtuality ($t^2\gg \hat{q}E$) regime is described by the Dokshitzer-Gribov-Lipatov-Altarelli-Parisi (DGLAP) evolution modified to include nuclear medium effects \cite{Majumder:2011uk,Majumder:2009zu,Wang:2009qb} based on the higher-twist formalism \cite{Wang:2001ifa,Majumder:2009ge,Qin:2009gw}. The jet-medium interaction is encapsulated in various transport coefficients governing the exchange of energy-momentum between jet partons and those in the QGP: The transverse momentum diffusion of jet partons in the QGP is encoded in $\hat{q}$~\cite{Baier:2002tc,Kumar:2020wvb}, while longitudinal transfers (not used herein) are contained within $\hat{e}$ and $\hat{e}_2$~\cite{Peshier:2006hi,Majumder:2008zg}. 

With every split in the DGLAP or vacuum like stage the virtuality undergoes a reduction. Once the virtuality reaches $t_s\sim\sqrt{\hat{q} E}$, the switching virtuality ($t_s$) between the DGLAP and transport stages, multiple scatterings from the medium maintain the virtuality at the scale $\sqrt{\hat{q}E}$. In this effort, $t_s$ is treated as a free parameter, tuned using Bayesian methods. Below $t_s$, the virtuality scale is considered close to that of the medium in our simulations, and thus rate equations \cite{Jeon:2003gi,Turbide:2005fk,Qin:2009bk} become an apt description of parton evolution in the QGP. Finally, once partons reach low energies and (and low virtualities), hadronization occurs via PYTHIA's string fragmentation present within the JETSCAPE framework~\cite{Putschke:2019yrg,JETSCAPE3.5}. The high and low virtuality parton energy loss regimes are incorporated inside the JETSCAPE framework, which provides a model-agnostic communication layer among jet energy loss models, allowing for a multi-stage event generator to be created, as is the case in this study. Beyond jet-medium interactions, the model-agnostic nature of the JETSCAPE framework has been used to study and interpret simulations of the nuclear bulk medium itself~\cite{JETSCAPE:2020mzn,JETSCAPE:2020shq}, while the framework also provides a dynamical communication layer between simulation of the nuclear medium and the energy-loss calculations. Finally, the JETSCAPE framework has developed a set of Bayesian tools to constrain jet-quenching calculations in heavy-ion collisions~\cite{JETSCAPE:2021ehl}. 

This study focuses on improving Bayesian tools to constrain the nuclear modification factor of inclusive jets, light hadrons and D-mesons (a discussion on particulars of heavy flavor production can be found in Ref.~\cite{Andronic:2015wma}). The combination of highly dimensional parameter spaces explored in high-energy heavy-ion collisions simulations (e.g.~\cite{Bernhard:2019bmu,JETSCAPE:2020mzn,Nijs:2020roc,JETSCAPE:2021ehl,Heffernan:2023utr,Liyanage:2023nds,Liu:2023rfi}) together with the high computational requirements to generate realistic simulations, necessitates the use of model emulators to accelerate the Markov Chain Monte Carlo computations employed when obtaining the posterior parameter distributions in Bayesian analysis. Given that the presence of emulators is currently unavoidable inside large-scale Bayesian analysis, a new measure quantifying the performance of an emulator is needed.

For reliable predictions with quantified uncertainty, we introduce herein a new measure for quantifying the performance of Gaussain Process (GP) emulators to approximate full model calculations. This novel measure is based on the Kullback-Leibler divergence within closure tests.  The novel measure proposed is inspired from the work done in devising scoring rules \cite{Matheson:1976,Winkler:1994,Gneiting:2007,Dawid:2014} in Bayesian statistics. To fully appreciate the usefulness of this new measure, a Bayesian inference (calibration) of model parameters is performed where a selection of hadron and inclusive jet suppression observables are used, focusing solely on central ($0-10\%$) nuclear collisions at the LHC. As our calculation currently yields a large statistical uncertainty for heavy flavor observables, validation of the Bayesian inference workflow is especially important, and presents a good scenario to test our new performance metric of GP emulators. We also test its sensitivity to various model parameters and to different observables, thus highlighting its usefulness for future Bayesian analysis. Finally, we present Bayesian constraints on model parameters explored in our limited Bayesian analysis, and provide comparisons with experimental data. 

This work is organized as follows: Sec.~\ref{sec:sim_setup} presents details regarding the multi-stage energy loss calculation herein as well as provide details about the hydrodynamical simulation of the QGP with which partons will interact. Section~\ref{sec:bayesian inference} presents our Bayesian setup, with the new measure quantifying the performance for GP emulation being presented in Sec.~\ref{sec:quantitative_closure_test}. The best performing GP emulator is then used within a small-scale Bayesian calibration. Section~\ref{sec:conclusion} is reserved for concluding remarks and present an outlook of how our current Bayesian analysis can be improved in the future. 
\section{Simulation Setup}
\label{sec:sim_setup}

In the following sections we describe our simulation of parton evolution in the QGP. Section \ref{sec:energy_loss} describes the the models used to explain the interaction of jet partons with the QGP, while Sec. \ref{sec:medium_evol} provides details of how QGP is evolved. 
\subsection{Parton interactions with the QGP}
\label{sec:energy_loss}
After initial parton production in PYTHIA, the evolution of high-energy and high-virtuality partons is calculated in MATTER (Modular All Twist Transverse-scattering Elastic-drag and Radiation) \cite{Majumder:2013re,Cao:2017qpx}, which uses the higher twist formalism \cite{Wang:2001ifa,Majumder:2009ge,Qin:2009gw} to explain parton interactions within the QGP. A virtuality ordered shower is thus developed for massless and massive \cite{Abir:2015hta,JETSCAPE:2022hcb} partons. Once hard partons in the shower reach a low virtuality regime, further evolution proceeds via the Linear Boltzmann Transport (LBT) model \cite{Luo:2018pto}. The LBT interactions between the hard partons and the QGP are preserving parton virtuality while modifying their energy, and three-momentum direction. Thus, MATTER evolves partons with virtuality $t > t_s$ --- $t_s$ being the switching virtuality --- while LBT simulates those with $t\leq t_s$. The connection time between the PYTHIA shower and the energy loss models is chosen to be $0.6$ fm/$c$ but the dependence of the nuclear modification factor on this quantity is found to be weak \cite{JETSCAPE:2022jer,JETSCAPE:2023hqn}. Following the evolution in LBT, the JETSCAPE framework determines whether partons undergo further splittings in MATTER (i.e. for parton leaving the QGP with enough virtuality to continue showering in the vacuum) or whether they need to hadronize (hadronization is handled via fragmentation in PYTHIA) using the colorless string hadronization routine \cite{JETSCAPE:2019udz}.
\subsubsection{The MATTER simulation}
\label{sec:MATTER}
Parton decays in MATTER are calculated using the Sudakov form factor. The probability for {\it no} decay for a parton is given by:
\begin{eqnarray}
\Delta\left(t,t_{\rm min}\right) &=& \exp\left[-\int^t_{t_{\rm min}} dt' \int^{z_{\rm max}}_{z_{\rm min}} dz \frac{dN^{\rm tot}}{dz dt'}\right]\nonumber\\
\frac{dN^{\rm tot}}{dz dt'} &=& \frac{dN^{\rm vac}}{dz dt'} + \frac{dN^{\rm med}}{dz dt'},
\label{eq:sud_Delta}
\end{eqnarray} 
where $\frac{dN^{\rm vac}}{dz dt'} + \frac{dN^{\rm med}}{dz dt'}$ includes all possible decay channels of a given parton. For instance, according to soft collinear effective theory (SCET) \cite{Abir:2015hta}, the decay of heavy quark $Q\to Q+g$ gives:
\begin{widetext}
\begin{eqnarray}
\frac{dN^{\rm vac}}{dz dt} + \frac{dN^{\rm med}}{dz dt } &=& \frac{\alpha_s(t)}{2\pi} \frac{P_{g\gets Q}(z)}{t}\left\{ 1+ \int^{\tau^+_Q}_0 d\tau^+\frac{2-2\cos\left( \frac{\tau^+}{\tau^+_Q} \right)}{z(1-z)t(1+\chi)^2} \left[ \left(\frac{1+z}{2}\right) - \chi + \left(\frac{1+z}{2}\right) \chi^2 \right]\hat{q}  \right\}\nonumber\\
\label{eq:hq_ht}
\end{eqnarray} 
\end{widetext}
In Eq.~(\ref{eq:hq_ht}), $z$ labels the momentum fraction of the daughter heavy quark, $M$ is the mass of the heavy quark, $\chi=(1-z)^2M^2/l^2_\perp$, with $l^2_\perp$ being the relative transverse momentum square between the outgoing daughter partons, determined via $z(1-z)t=l^2_\perp(1+\chi)$, while $t$ is the virtuality of the heavy quark and $P_{g\gets Q}(z)=C_F\left(\frac{1+z^2}{1-z^{\,\,\,}}\right)$ is the splitting function and $C_F=4/3$. The light flavor result \cite{Wang:2001ifa,Majumder:2009ge,Qin:2009gw,Majumder:2013re}, is recovered in the limit $M\to 0$. The integral over light-cone time $\tau^+$ in Eq.~(\ref{eq:hq_ht}) assumes the medium is in its rest frame, with the upper bound $\tau^+_Q=2q^{+}/t$ being given by the ratio of forward light-cone momentum $q^+=\left(q^0+{\bf q}\cdot \hat{n}\right)/\sqrt{2}$ (with $\hat{n}={\bf q}/\vert {\bf q}\vert$), and the virtuality $t$. 

As the $g\to Q+\bar{Q}$ has not yet been calculated using the SCET approach devised in Ref.~\cite{Abir:2015hta}, this phenomenological study estimates the gluon splitting into heavy quarks using the light flavor formula \cite{Majumder:2013re}, and reduces the kinematic range using \cite{JETSCAPE:2022hcb}
\begin{eqnarray}
z_{\rm min} &=& \frac{t_0+M^2}{t}+\mathcal{O}\left(\left(\frac{t_0+M^2}{t}\right)^2\right)\nonumber\\
z_{\rm max} &=& 1-\frac{t_0+M^2}{t}+\mathcal{O}\left(\left(\frac{t_0+M^2}{t}\right)^2\right),
\end{eqnarray} 
assuming $M^2/t\ll 1$, $t_0/t\ll 1$, and $t_0=1$ GeV$^2$. Imposing $z_{\rm max} > z_{\rm min}$ as well as $t>t_{\rm min}$, requires that $t_{\rm min}=2(M^2+t_0)$. The determination of $t$ and $z$ proceeds in the same way as for $Q\to Q+g$ (more details are in Ref.~\cite{JETSCAPE:2022hcb}). 

The transverse momentum broadening $\hat{q}[T(\tau^+)]$ acquired by the quark as it traverses the QGP is the only quantity that depends on $\tau^+$ through the temperature $T$. From Hard Thermal Loop (HTL) approximation~\cite{He:2015pra}, $\hat{q}$ is 
\begin{eqnarray}
\hat{q}^{HTL}=C_a \frac{42\zeta(3)}{\pi}\alpha_s^2T^3 \ln \left(\frac{cET}{4m^2_D}\right)
\end{eqnarray}
where $\zeta(3)\approx 1.20205$ is Ap\'{e}ry's constant, $C_a=N_c=3$ the number of colors, while the Debye mass is $m^2_D=6\pi\alpha_s T^2$, and $c \approx 5.7$ \cite{Caron-Huot:2009fku}. The studies \cite{JETSCAPE:2022jer,JETSCAPE:2022hcb} showed that a constant effective $\alpha^{\rm eff}_s$ can be improved by allowing the coupling to run with the scale $\mu^2=2ET$ via
\begin{eqnarray}
\hat{q}^{HTL}=C_a \frac{42\zeta(3)}{\pi}\alpha_s(\mu^2) \alpha^{\rm eff}_s T^3 \ln \left(\frac{cET}{4m^2_D}\right)
\label{eq:qhat_HTL}
\end{eqnarray}    
where
\begin{eqnarray}
m^2_D&=&\frac{4\pi\alpha^{\rm eff}_s T^2}{3} \left(N_c+\frac{N_f}{2}\right) \overset{N_f=3}{=}6\pi\nonumber\alpha^{\rm eff}_s T^2\\
\alpha_s(\mu^2)&=&\left\{
\begin{array}{rl}
\alpha^{\rm eff}_s & \mu^2 < \mu^2_0, \\
\frac{4\pi}{11-2N_f/3} \frac{1}{\ln\frac{\mu^2}{\Lambda^2}}& \mu^2 > \mu^2_0,\\
\end{array} \right.
\end{eqnarray}
with $\Lambda$ being chosen such that $\alpha_s\left(\mu^2_0\right)=\alpha^{\rm eff}_s$ at $\mu^2_0=1$ GeV$^2$ \cite{Kumar:2019uvu}. The effective parametrization of the $t$-dependent $\hat{q}$ is \cite{JETSCAPE:2022jer,JETSCAPE:2022hcb}
\begin{eqnarray}
\frac{\hat{q}(t)}{\hat{q}^{HTL}}=H(t)=\frac{c_0}{1+c_1\ln^2(t)+c_2\ln^4(t)}
\label{eq:qhat_t}
\end{eqnarray}
where $\hat{q}^{HTL}$ is given in Eq.~(\ref{eq:qhat_HTL}), $c_1$ as well as $c_2$ are tunable parameters, and $c_0=1+c_1\ln^2(t_s)+c_2\ln^4(t_s)$ is an overall normalization ensuring $\frac{\hat{q}(t)}{\hat{q}^{HTL}}\in [0,1]$ for $t>t_s$. Note that currently the virtuality dependence of $\hat{q}$ is assumed to be the same regardless of the mass of the quark \cite{JETSCAPE:2022hcb}. Finally, MATTER also includes elastic $2\to 2$ scatterings using leading order perturbative QCD matrix elements as explored in detail below. 

\subsubsection{The linearized Boltzmann transport simulation}
\label{sec:LBT}
Once a parton enters the linearized Boltzmann transport (LBT) at $t<t_s$, its virtuality remains unchanged (see e.g. \cite{Cao:2020wlm} and references therein). The LBT relies on solving the Boltzmann equation taking into account $2\to 2$ and $2\to 3$ processes. The $2\to 2$ scattering processes consist of leading order perturbative QCD matrix elements. The evolution of the momentum and position distribution of a hard quark $Q$ with momentum $p_1$ is given by:
\begin{widetext}
\begin{eqnarray}
p^\mu_1 \partial_\mu f_1(x_1,p_1)&=&\mathcal{C}_{\rm el}[f_1]+\mathcal{C}_{\rm inel}[f_1]\nonumber\\
\mathcal{C}_{\rm el}[f_1] &=& \frac{d_2}{2}\int dP_2 \int dP_3 \int dP_3 (2\pi)^4\delta^{(4)}\left(p_1+p_2-p_3-p_4\right)\left\vert \mathcal{M}_{1,2\to3,4}\right\vert^2 \lambda_2\left(s,t,u\right)\times \nonumber\\
&\times& \left\{f_3\left({\bf p}_3\right)f_4\left({\bf p}_4\right)\left[1\pm f_1\left({\bf p}_1\right)\right]\left[1\pm f_2\left({\bf p}_2\right)\right]-f_1\left({\bf p}_1\right)f_2\left({\bf p}_2\right)\left[1\pm f_3\left({\bf p}_3\right)\right]\left[1\pm f_4\left({\bf p}_4\right)\right]\right\}
\end{eqnarray}
\end{widetext}
where $d_2$ is the spin-color degeneracy of parton ``2'', $\int dP_i \equiv \int \frac{d^3 p_i}{(2\pi)^3 2 p^0_i}$ with $i=2,3,4$; while $\lambda_2\left(s,t,u\right)=\theta\left(s-2m^2_D\right) \theta\left(s+t-m^2_D\right)\theta\left(-t-m^2_D\right)$. The same $2\to 2$ scattering rates are also used in MATTER. 

The medium-induced gluon radiation describing $2\to 3$ processes uses the same higher twist formulation as that employed in Eq.~(\ref{eq:hq_ht}) of the MATTER simulation. The latter has an average number of gluons emitted from a hard quark (between time $t$ and $t + \Delta t$):
\begin{eqnarray}
\bar{N}^{\rm med}(t\to t + \Delta t) \approx\Delta t\int dz dk_{\perp}^2 \frac{dN^{\rm med}}{dz dk_{\perp}^2 dt}\nonumber\\
\frac{dN^{\rm med}}{dz dk_{\perp}^2 dt}= \frac{2\alpha_s P(z)}{\pi k_{\perp}^4}\hat{q}\left(\frac{k_{\perp}^2}{k_{\perp}^2+z^2M^2}\right)^4\sin^2\left(\frac{t-t_i}{2\tau_f}\right).\nonumber\\
\end{eqnarray}
A Poisson probability distribution is employed to sample independent successive emissions, with the probability of emitting $n$ gluons being
\begin{equation}
\mathcal{P}(n)=\frac{\left(\bar{N}^{\rm med}\right)^{n}}{n!}e^{-\bar{N}^{\rm med}},
\end{equation}
while the probability of a total inelastic process is $\mathcal{P}_{\rm inel.}=1-e^{-\bar{N}^{\rm med}}$. The procedure to determine whether (and how many) elastic vs inelastic scatterings inside the QGP have occurred is explored in detail in Ref. \cite{JETSCAPE:2022jer}. The only undetermined coefficient in LBT is the strong coupling $\alpha_s$, which has a fixed component $\alpha^{({\rm eff})}_s=0.3$, and a running component $\alpha_s(\mu^2)$ \cite{JETSCAPE:2022hcb}.

\subsection{Evolution of the QCD medium}
\label{sec:medium_evol}
The evolution of the QCD medium used herein is performed using a boost-invariant 2+1-dimensional model which involves three stages: a pre-hydrodynamic, hydrodynamic, and a hadronic transport stage~\cite{Bernhard:2019bmu,Shen:2014vra,Bass:1998ca,Bleicher:1999xi}. The pre-hydrodynamic stage is based on the \trento{} initial condition for Pb-Pb collisions \cite{Moreland:2014oya}, which is followed by a collisionless Boltzmann evolution for a proper time of $\tau_{FS}=1.2$ fm/$c$. Free-streaming generates a non-trivial initial profile used inside a 2+1D hydrodynamical simulation. 400 \trento{} initial Pb-Pb configurations were generated within the 0-10\% centrality class at $\sqrt{s_{NN}}=5.02$ TeV (for more details see Ref.~\cite{Fan:2022ton}). The relevant parameters used for simulating the evolution of the QCD medium are extracted from a Bayesian model-to-data comparison, explained in \cite{Bernhard:2019bmu,Bernhard:2018hnz}. The event-by-event setup of the soft medium has been found to be important for the proper description of jet energy-loss due to the added fluctuations in the medium \cite{Noronha-Hostler:2016eow}. The hydrodynamical simulation \cite{Song:2007ux,Shen:2014vra} is stopped once all fluid cells reach below $T_c=154$ MeV \cite{Bazavov:2014pvz}, at which point all fields are converted into particles using the Cooper-Frye prescription \cite{McNelis:2019auj,Bernhard:2018hnz,Huovinen:2012is}, following which the Ultrarelativstic Molecular Dynamics (UrQMD) \cite{Bass:1998ca,Bleicher:1999xi} hadronic (Boltzmann) transport simulation is used.

\section{Bayesian Inference}\label{sec:bayesian inference}
There are four parameters in the aforementioned multistage energy loss approach: the effective coupling constant $\alpha^{\rm eff}_s$, the switching virtuality $t_s=Q_s^2$, and $(c_1,c_2)$ that control the virtuality dependence of $\hat{q}(t)$. Previous studies \cite{JETSCAPE:2022jer,JETSCAPE:2022hcb} have briefly explored the effects of these parameters on the charged hadron, D-meson, and inclusive jet $R_{AA}$. However, the full correlation between the parameters and the observables, as well as the full capability of this multistage approach to describe experimental data, remain to be quantified. Bayesian inference can help answer these questions. 

The Bayes' theorem states that the posterior distribution of the parameter set $\mathbf{x}$, given the experimental observation $\mathbf{y}_{\rm exp}$, is proportional to the product of the prior distribution $q(\mathbf{x})$ and the likelihood function $\mathcal{L}(\mathbf{y}_{\rm exp}|\mathbf{x})$:
\begin{equation}
    p(\mathbf{x}|\mathbf{y}) \propto \mathcal{L}(\mathbf{y}_{\rm exp}|\mathbf{x})q(\mathbf{x}).
\end{equation}
The prior $q(\mathbf{x})$, as the name suggests, represents our prior knowledge of the parameter values. The likelihood function $\mathcal{L}(\mathbf{y_{\rm exp}}|\mathbf{x})$ is the probability of observing $\mathbf{y}_{\rm exp}$ given a specific parameter set $\mathbf{x}$:
\begin{equation}
\mathcal{L}(\mathbf{y}|\mathbf{x}) = \frac{\exp\left[-\frac{1}{2} \left[\mathbf{f(\mathbf{x})} - \mathbf{y}_{\rm exp}\right]^{\intercal} \Sigma^{-1} \left[\mathbf{f(\mathbf{x})} - \mathbf{y}_{\rm exp}\right]\right]}{\sqrt{(2\pi)^m \rm{det} \Sigma}} ,
\end{equation}
where $m$ is the dimension of $\mathbf{y}_{\rm exp}$, $\Sigma = \Sigma_{M} + \Sigma_{\rm exp}$ is the uncertainty covariance matrix, which takes into account both model and experimental uncertainties, and $f(\mathbf{x})$ is the model calculation given the parameters $\mathbf{x}$. In the case of model uncertainties $\Sigma_M$, solely statistical are accounted for herein.

If $f(\mathbf{x})$ is known for an arbitrary $\mathbf{x}$, then one can perform a Markov chain Monte Carlo (MCMC) random walk through the parameter space to extract the posterior parameter distribution. However, each point in the parameter space requires at least $\mathcal{O}(10^4)$ CPU hours to compute, meaning it is computationally prohibitive to walk in this parameter space by performing a full model simulation at each step. A surrogate model that can mimic the actual model with a reasonable computational cost is needed. The Gaussian Process (GP) emulator is chosen as a fast surrogate model yielding both mean and covariance information. The surrogate model is trained on the set of pre-computed $(\mathbf{x},f(\mathbf{x}))$ pairs called the training data stored as $(X_{\rm train},\mathbf{y}_{\rm train})$. $X_{\rm train}$ has dimensions $m\times k$ where $m$ is the number of training data and $k$ is the dimension of the parameter set. $\mathbf{y}_{\rm train}$ is a $m \times 1$ vector, since at each training point, just one dimension of the model output $f(\mathbf{x})$ is emulated. A GP essentially interpolates between all training data. Mathematically, one assumes that all desired outputs $\mathbf{y}$ to be predicted at inputs $X$, along with the known outputs $\mathbf{y}_{\rm train}$ at the training points $X_{\rm train}$, follow a multivariate normal distribution:
\begin{align}
& \begin{pmatrix}
\mathbf{y} \\
\mathbf{y}_{\rm train} 
\end{pmatrix} \label{eqn:GP_predict} \\
& \sim 
\mathcal{N}  \left( \begin{pmatrix}
\mu\\
\mu_{\rm train}
\end{pmatrix}, 
\begin{pmatrix}
K(X,X) & K(X, X_{\rm train}) \\
K(X_{\rm train}, X) & K(X_{\rm train},X_{\rm train})
\end{pmatrix} \right), \nonumber
\end{align}
where $K$ denotes the covariance matrix. The distribution of $\mathbf{y}$ is then given by:
\begin{widetext}
\begin{equation}
    \begin{split}
    \mathbf{y} \sim  &~\mathcal{N}(K(X, X_{\rm train})K^{-1}(X_{\rm train},X_{\rm train})\mathbf{y}_{\rm train},  K(X,X)-K(X, X_{\rm train})K^{-1}(X_{\rm train},X_{\rm train}) K(X_{\rm train}, X)).
    \end{split}
\end{equation}
\end{widetext}
Each element in the covariance matrix $K$ is calculated with the kernel function $k(\mathbf{x},\mathbf{x}')$ that characterizes the correlation between two points in the parameter space \cite{JETSCAPE:2020mzn}. The kernel encodes the prior Bayesian belief of the function being mimicked. Common kernel choices include \cite{williams2006gaussian}:
\begin{enumerate}
    \item the radial basis function (RBF) kernel:\\ $k(r)=\sigma^2\exp\left(-\frac{r^2}{2l^2}\right)$ (equivalent to the Mat\'{e}rn kernel with $\nu \rightarrow \infty$).
    \item the Mat\'{e}rn ($\nu=5/2$) kernel:\\ $k(r)=\sigma^2\left(1+\frac{\sqrt{5}r}{l}+\frac{5r^2}{3l^2}\right)\exp\left(-\frac{\sqrt{5}r}{l}\right).$
    \item the Mat\'{e}rn ($\nu=3/2$) kernel:\\ $k(r)=\sigma^2\left(1+\frac{\sqrt{3}r}{l}\right)\exp\left(-\frac{\sqrt{3}r}{l}\right).$
    \item the white noise kernel: $k(r)=\sigma^2\delta(r).$
\end{enumerate}
where $r=|x-x'|$, while $\sigma$ and $l$ are hyperparameters that are assigned a possible window and then optimized to maximize the likelihood of fit of the Gaussian process to the training data.

Since a GP emulator maps onto a one-dimensional space, in principle one would need $\rm dim(\mathbf{y})$ of GP emulators for all the data points. However, as the simulation results are in fact correlated (e.g., the measured/calculated $R_{AA}$ points at different $p_T$ are positively correlated with each other), dimensional reduction of the data is possible using principal component analysis (PCA). The principal component decomposition allows to select the number ($N_{\rm PC}$) of principal components (PC) --- a subset of vectors in $\rm dim(\mathbf{y})$-dimensional space --- that emulate the majority of the variance in training data. The additional uncertainty introduced by GP emulation, and truncating uncertainty induced via PCA selection,
are accounted for in the covariance matrix $\Sigma$.

The last ingredient needed for an efficient Bayesian inference workflow is to devise the optimal distribution of training data points. Latin hypercube sampling of the parameter space is used to optimally distribute training points. One runs the full simulations at the hypercube-sampled design points, selects the first few principal components in PCA ($N_{\rm PC}$ containing the best optimal number of PCs) , and trains $N_{\rm PC}$ GP emulators using the appropriate kernel functions discussed above. To predict the model output at a new point in the parameter space, one runs the $N_{\rm PC}$ GP emulators at this new point and then inverse-transforms the outputs from the principal component space onto the original observable (output) space. Besides generating the training data, the two key factors that affect model emulation process is the choice of the kernel in GP, and the number of principal components $N_{\rm PC}$ in PCA. One of the main results of this work --- presented in Sec.~\ref{sec:quantitative_closure_test} --- is to devise a quantitative measure of the performance of the GP emulator, allowing to choose the best emulator for a Bayesian model-to-data compoarison. Finally, using the optimal emulator, the MCMC random walk in the parameter space is performed to extract the posterior distribution of the parameters within a limited Bayesian model-to-data comparison.

\subsection{Calibration setup}
The prior range for the model parameters are considered to be uniform distributions and are listed in Table~\ref{table-1}.
\begin{table}[h!]
\centering
\caption{\label{table-1} Prior parameters ranges in our Bayesian calibration.}
\begin{tabular}{ | c | c | c | c | c | c | c | c | c}
 \hline
 Parameter & $\alpha^{\rm eff}_s$ &  $Q_s$  & $c_1$  & $c_2$ \\
 \hline
 Range     & 0.1 - 0.5  & 1.5 - 4 & 1 - 10 & 50 - 300\\
 \hline
 \end{tabular}
\end{table}
\begin{figure}
	\centering
	\includegraphics[width=0.5\textwidth]{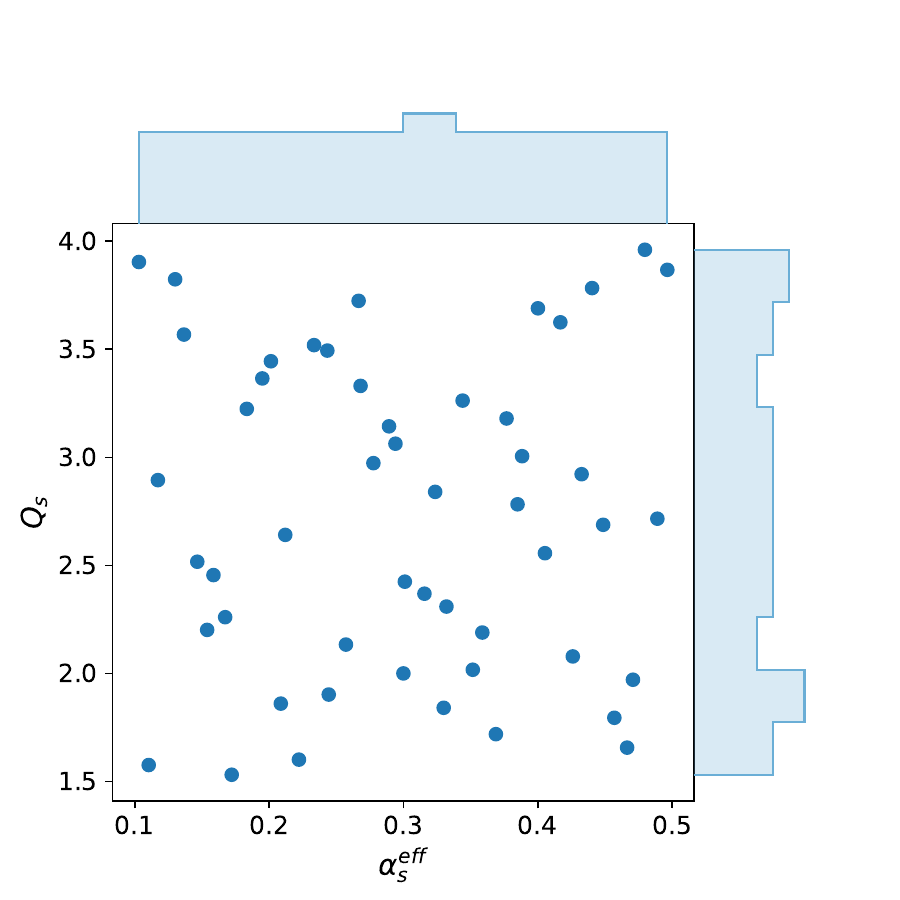}
        \vspace{-0.75cm}
	\caption{ Distribution of input parameter $\alpha^{\rm eff}_s$ and $Q_s$ from all 50 sampled design points within the prior range.}
\label{fig:Design14}
\end{figure}
Those ranges are selected based on previous exploration of these parameters in Ref.~\cite{JETSCAPE:2022jer,Fan:2022ton}. Due to constrained computation budget of this work, 50 design points are drawn by Latin hypercube sampling (see Fig.~\ref{fig:Design14}). For each design point, roughly 400,000 events are generated and distributed evenly among 400 fluid simulations. Sizable statistical fluctuations are observed especially for charged hadron and D-meson $R_{AA}$, which will impact our calibration. The validity of the Bayesian analysis against model calculation fluctuations will be verified in Sec.~\ref{sec:quantitative_closure_test}, with further details in Appendix \ref{appdx:1}.

Given the computational resources available, this work focuses on Pb-Pb collision at $\sqrt{s_{NN}}= 5.02$~TeV and 0-10\% centrality. A previous study \cite{Xu:2017obm} shows that calibrating to different collision energies independently versus at the same time may yield slightly different posteriors. Equivalently, one may need to use different values for the same parameter in different collision systems \cite{Ke:2020clc}. We leave such a exploration to a future study. 
\begin{figure*}
    \centering
    \includegraphics[width=\textwidth]{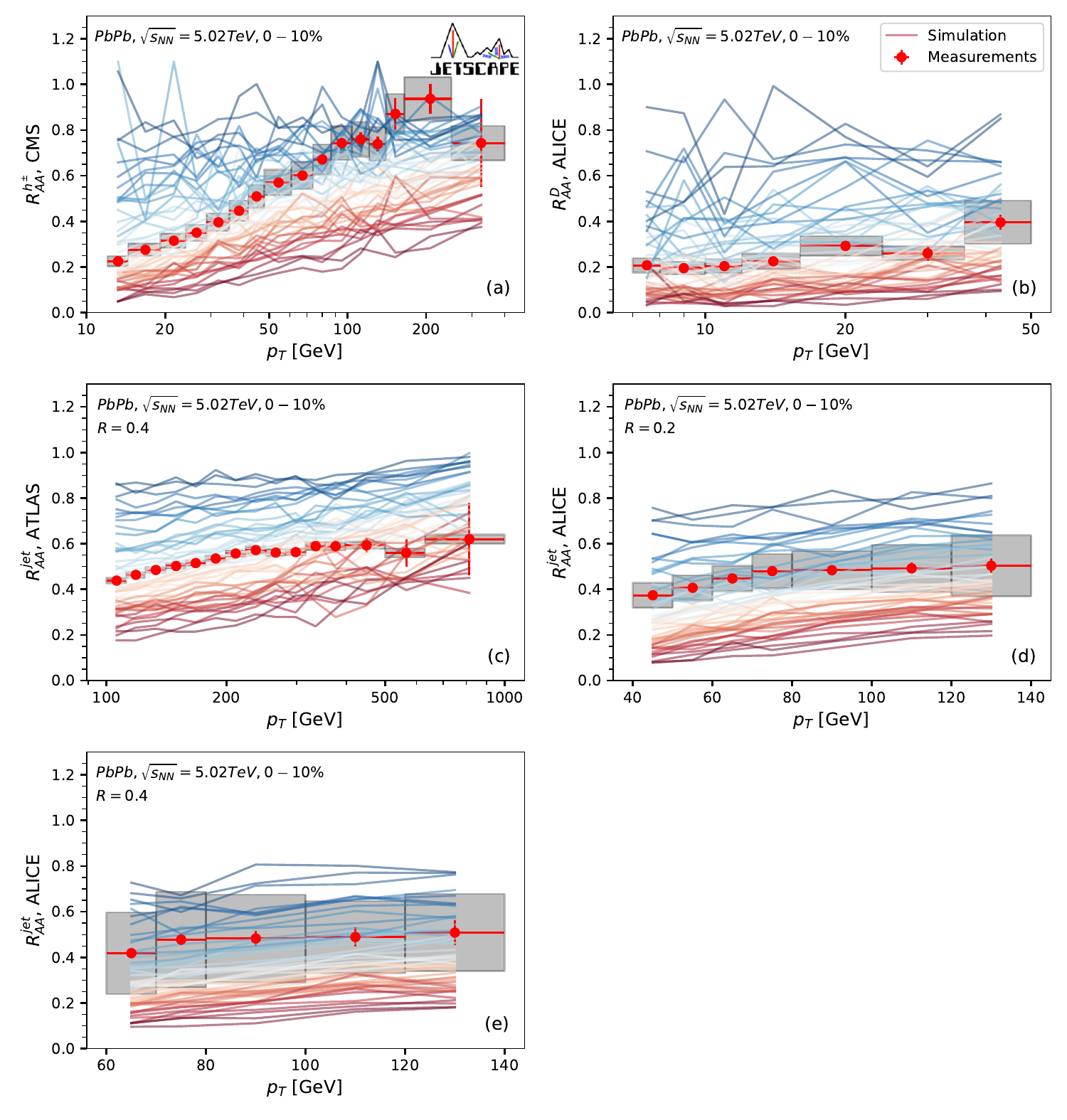}
    \vspace{-0.75cm}
    \caption{(Color online) A comparison between model calculation using parameters from all the design points and experimental data is shown, focusing on LHC data from Pb-Pb collisions at $\sqrt{s}_{NN}=5.02$~TeV and 0-10\% centrality. Leading hadrons theory to data comparisons are presented in panels (a) and (b). Specifically, panel (a) shows charged hadron $R_{AA}$ theoretical calculations againts experimental data from CMS \cite{CMS:2016xef}, while in (b) focuses on D-meson $R_{AA}$ calculations against measurements from ALICE \cite{ALICE:2018lyv}. Panels (c) through (e) are reserved  inclusive jet $R_{AA}$ comparisons, where jets are reconstructed using the anti-${\rm k_T}$ algorithm. In (c) theoretical calculations are contrasted against data from ATLAS \cite{ATLAS:2018gwx}, while (d) and (e) centers on describing the jet radius dependence using data from ALICE \cite{ALICE:2019qyj} at $R=0.2$ and $R=0.4$, respectively. Each unique color in each plot corresponds to calculation from a single design point.}
\label{fig:PredictedDesign} 
\end{figure*}

As for the experimental observables that will be calibrated to, we choose the nuclear modification factor $R_{AA}$ for charged hadrons, D-mesons, and inclusive jets. $R^X_{AA}$ is defined as
\begin{eqnarray}
R^X_{AA}= \frac{\frac{d \sigma^X_{AA}}{d p_T}}{\frac{d \sigma^X_{pp}}{d p_T}}=\frac{\sum_{\ell} \frac{d N^X_{AA,\ell}}{d p_T}\hat{\sigma}_{\ell}(\hat{p}_T)}{\sum_{\ell} \frac{d N^X_{pp,\ell}}{d p_T}\hat{\sigma}_{\ell}(\hat{p}_T)}
\label{eq:R_AA}
\end{eqnarray}
where $\frac{dN^{X}_{AA}}{dp_T}$ and $\frac{dN^X_{pp}}{dp_T}$ are the multiplicity, in the experimentally given $p_T$ bin, of the quantity $X$, which is either charged hadrons, D-mesons, or jets originating from A-A and p-p collisions, respectively. The spectrum $\frac{d N^X_{AA,\ell}}{d p_T}$ is calculated utilizing the multistage model presented in Sec.~\ref{sec:energy_loss}. The total cross-section for producing a hard scattering process is broken down in several segment ($\ell$) of exchanged transverse momentum $\hat{p}_T$ contributing to the hard scattering at the level $\hat{\sigma}_{\ell}$. Each hard scattering event is sampled by PYTHIA. Many samplings of $\hat{\sigma}_{\ell}$, spanning a large kinematic range of the collision, are combined to produce $d\sigma^X/dp_T$. To avoid complications from hard-thermal hadronic recombinations and non-perturbative effects, the charged hadron and D-meson are sampled solely for $p_T\geq 7$~GeV.

An important step before training GP emulators is to verify that the dynamic range of model calculations spans that of the experimental observations. Our model calculations depicted in Fig.~\ref{fig:PredictedDesign} cover the measured range in $R^X_{AA}$, with each unique color in each plot corresponding to calculation from a single design point. 

At first glance, it may appear that the statistical fluctuations are significant, especially for D-meson $R_{AA}$, which could affect the validity of the Bayesian inference.\footnote{While generating the training data under a fixed computation budget, we were pursuing a balance between the number of design points to cover the parameter space and the number of events for each design point. This balance required a reduction in  statistics for the D-meson $R_{AA}$ at every design point, to ensure sufficient number of design points are available to cover the parameter space.} However, we have performed extensive validations of the GP emulator, in Sec.~\ref{sec:GPE_validation}, GP emulator closure tests in Sec.~\ref{sec:GPE_closure}, as well as stability tests of the posterior distribution in Appendix \ref{appdx:1} to addresses these concerns. 

\subsection{Emulator validation}
\label{sec:GPE_validation}
\begin{figure*}[htbp]
	\centering
	\includegraphics[width=\textwidth]{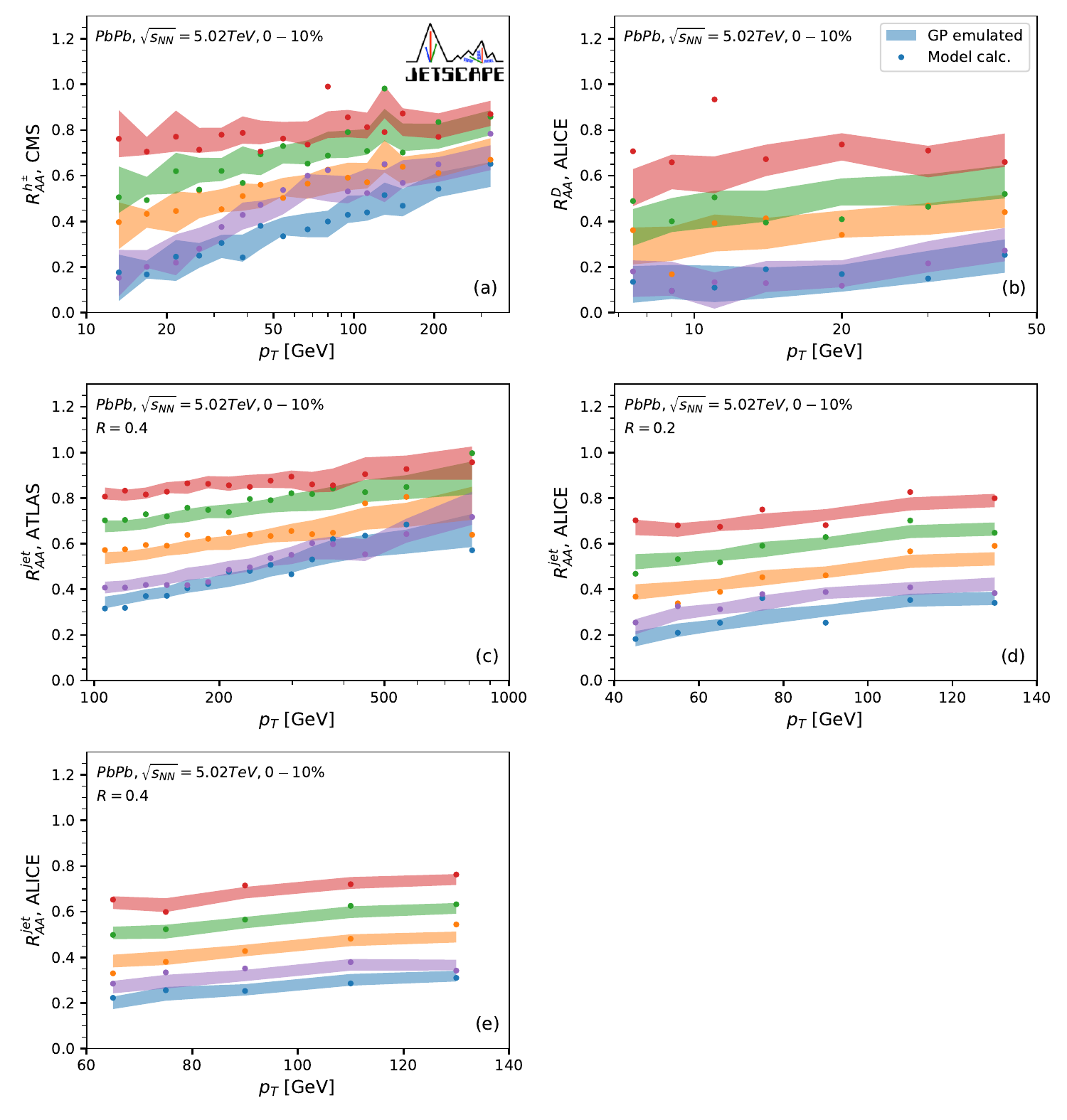}
        \vspace{-0.9cm}
        \caption{(Color online) Comparison between emulator predictions and model calculations at 5 random design points selected from the sample as the one depicted in Fig.~\ref{fig:Design14}. The combination of radial basis function with a white noise kernel is used along with 5 principal components (PCs). The panels follow the same categories as in    Fig.~\ref{fig:PredictedDesign}, while the colored bands correspond to the region covered by one standard deviation.}
\label{fig:EmulatorValidation} 
\end{figure*}
The GP emulator's performance validation is presented in Fig.~\ref{fig:EmulatorValidation}, where a direct comparison between emulator predictions and model calculations at $5$ random design points can be seen. Here the RBF with a white noise kernel is used along with 5 PCs. The emulator predictions fit the model calculations well, and seem to cut off some statistical fluctuations, stemming form using a subset of all PCs. 

\begin{figure*}
    \centering
    \vspace{-2cm}
    \includegraphics[width=\textwidth]{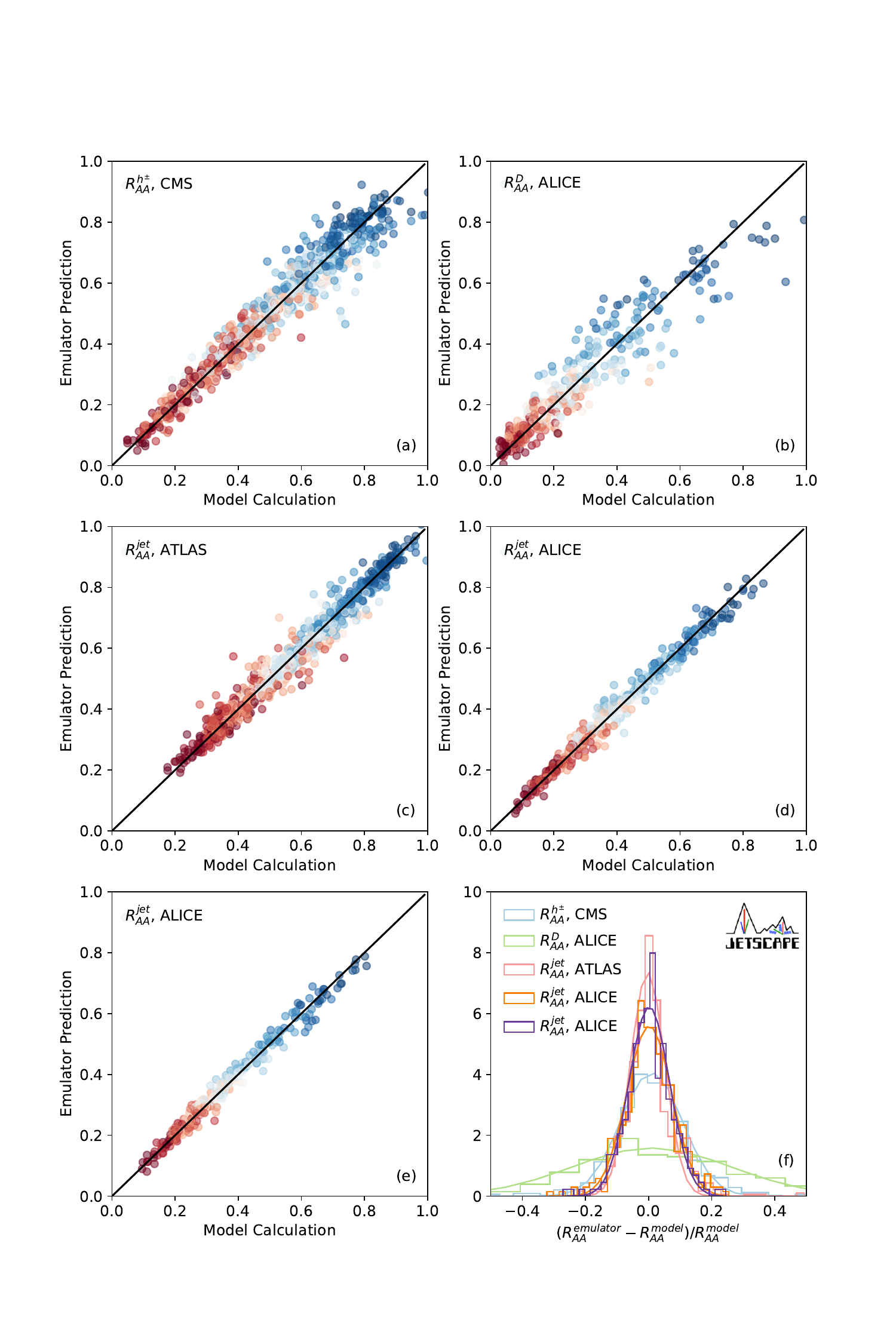}
    \vspace{-2.5cm}
    \caption{(Color online) A detailed comparison between emulator predictions and model calculations at all design points. The first five plots are scatter plots that display the model calculation and emulator  prediction for each observable. An ideal emulator gives the same output as the model, thus lying along the back solid line in panels (a) through (e). Each color in panels (a)--(e) are calculated using the same design point as in Fig.~\ref{fig:PredictedDesign}. Panel (f) shows histograms of the relative difference between model calculation and emulator prediction for different observables. The combination of RBF and white noise kernel, along with 5 PCs, are used throughout.}
\label{fig:EmulatorValidationScatter} 
\end{figure*}
The full performance of the emulator is plotted in Fig.~\ref{fig:EmulatorValidationScatter}, where the emulator response versus  model calculation is plotted for all design points. The emulator seems to perform the best at predicting inclusive jet $R_{AA}$ reconstricted using the anti-${\rm k_T}$ algorithm (see panels (c) -- (e)), followed by charged hadron $R_{AA}$ in panel (a), and finally the $D$-meson $R_{AA}$ in panel (d). Panel (f) of Fig.~\ref{fig:EmulatorValidationScatter} depicts the histograms of the relative difference between model calculations and emulator predictions for each observable. The distributions can all be fitted by a Gaussian centered near the origin, implying that the emulator induces little systematic bias in predicting the model. More quantitatively, the standard deviation is around $6-9\%$ when predicting inclusive jet $R_{AA}$, $12\%$ when predicting charged hadron $R_{AA}$, and $30\%$ when predicting D meson $R_{AA}$.\footnote{As a point of comparison, if all the design points are used to train a single emulator, the uncertainty of the relative difference will be slightly reduced (around $5-7\%$ when predicting inclusive jet $R_{AA}$, $10\%$ when predicting charged hadron $R_{AA}$, and $25\%$ when predicting D-meson $R_{AA}$). However we did not perform closure cross-validation on emulators using all design points, given the computational resources available for this work.} Note that when calculating the emulator prediction at one design point, the GP emulation training data set will exclude the data from that specific design point. Thus, different emulators are trained for each design point and the emulator does not know the truth values when making predictions, as required for cross-validation \citep{hastie2009elements} via closure tests. 
\subsection{Closure test}
\label{sec:GPE_closure}
Being able to predict the training data does not guarantee that our emulator can constrain the model parameters well. If the data are not sensitive to some parameters, those parameters may end up with a wide posterior. Furthermore, if there are degeneracies in the model, i.e. multiple combinations of model parameters can describe the same set of data, the posterior distributions becomes multimodal. These scenarios can be checked for by performing a cross-validation closure test, whereby one design point is taken out from the training process and treated as the truth. The emulator is trained without the truth point and then the posterior distributions of the parameters are drawn. Since the truth values for the parameters is known in this case, one can make a comparison between the posterior parameter distributions and the truth. If there are infinitely many design points and zero statistical fluctuation at each design point, one expects a very narrow peak in the posterior distribution in a closure test near the truth value (or several peaks in the case of degeneracy). 

\begin{figure*}
	\centering
	\includegraphics[width=0.325\textwidth]{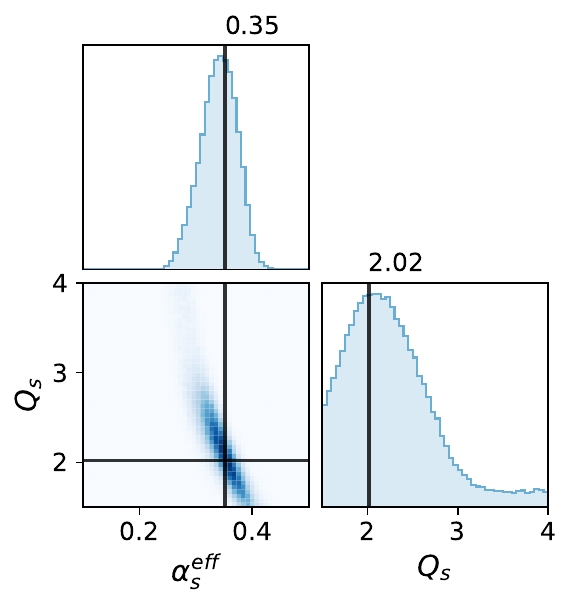}
	\includegraphics[width=0.325\textwidth]{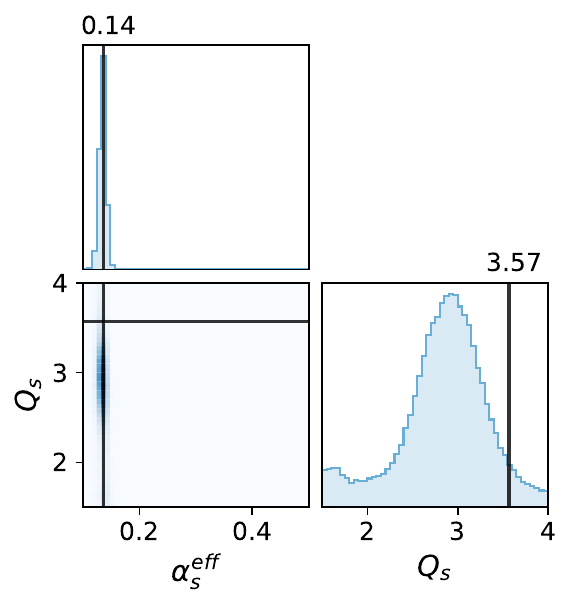}
	\includegraphics[width=0.325\textwidth]{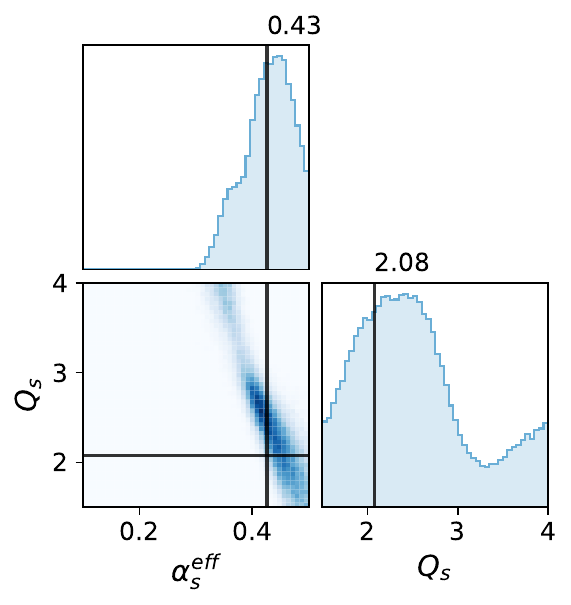}
	\includegraphics[width=0.325\textwidth]{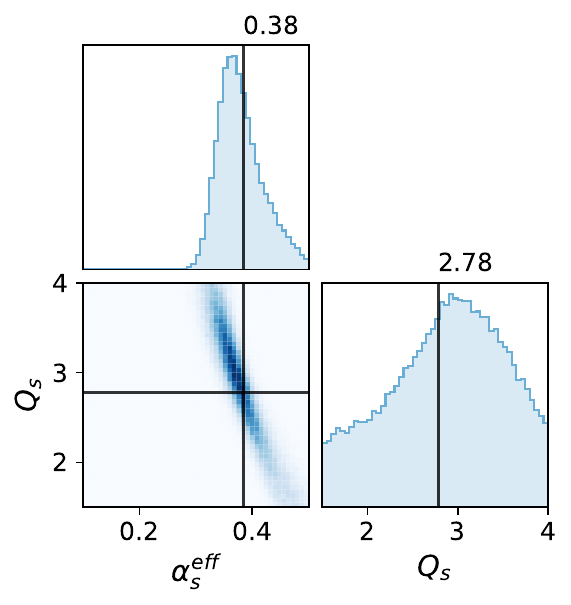}
	\includegraphics[width=0.325\textwidth]{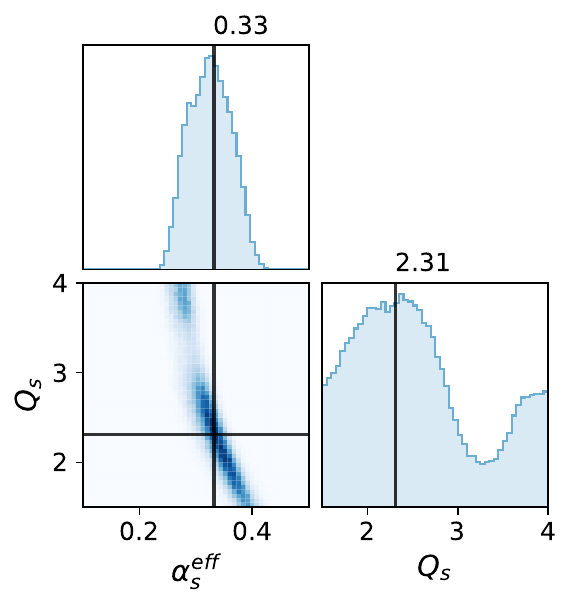}
	\includegraphics[width=0.325\textwidth]{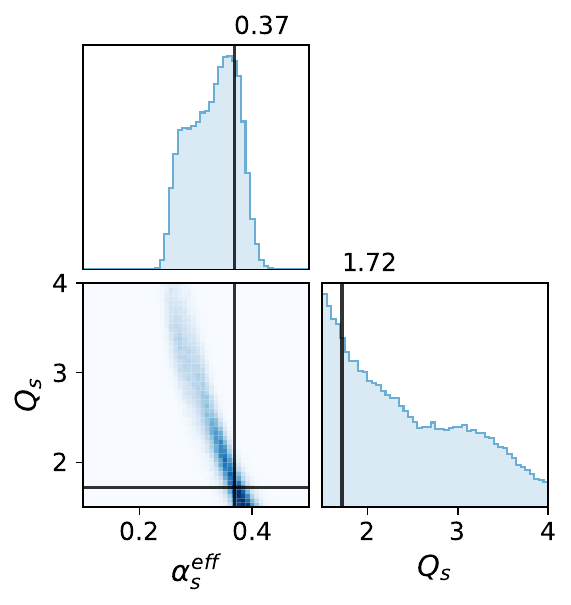}
	\includegraphics[width=0.325\textwidth]{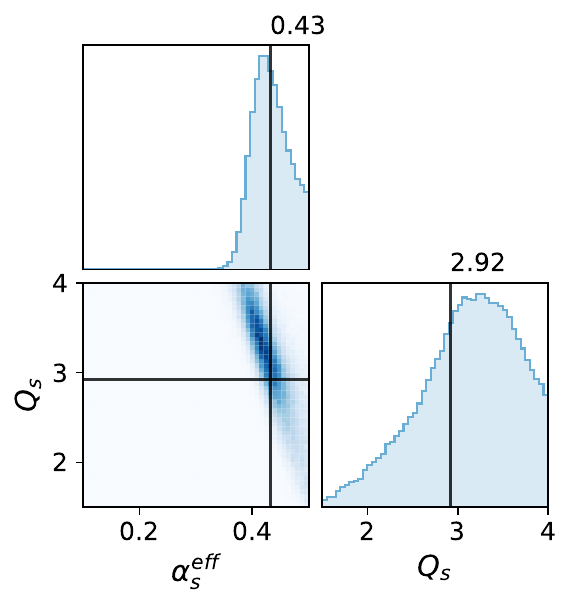}
	\includegraphics[width=0.325\textwidth]{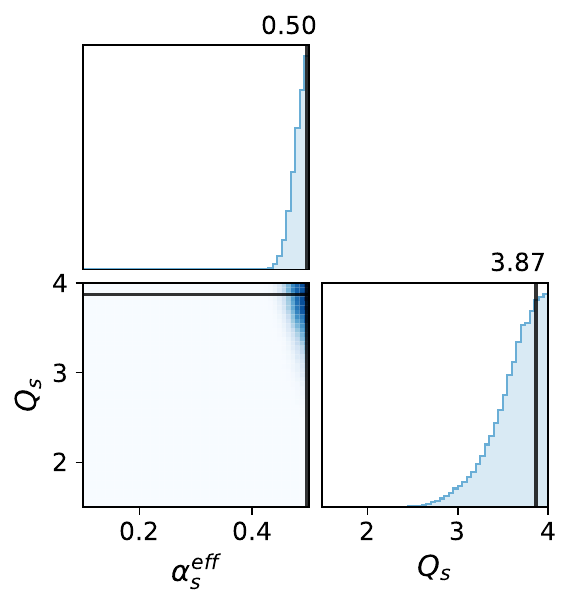}
	\includegraphics[width=0.325\textwidth]{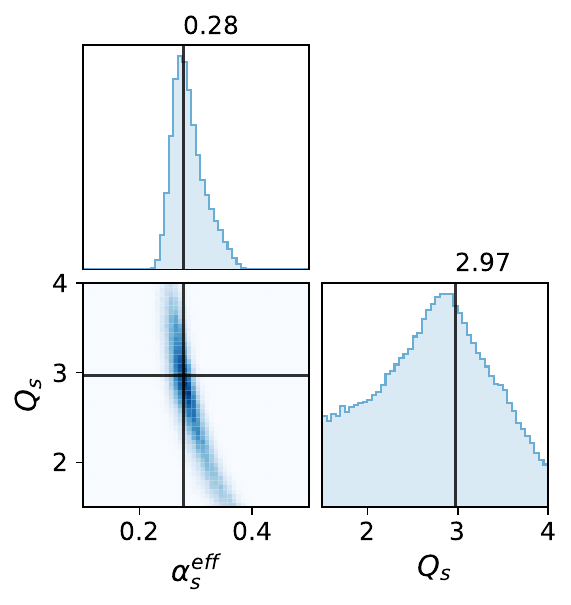}
        \vspace{-0.25cm}
	\caption{(Color online) Closure test results for $\alpha^{\rm eff}_s$ and $Q_s$ at $9$ random design points. The black lines represent the truth values. The RBF and white noise kernel is used along with 5 PCs.}
\label{fig:closure_exp_fluc_14} 
\end{figure*}

The results for the closure test at $9$ random design points is shown in Fig.~\ref{fig:closure_exp_fluc_14}. For each panel in this figure, a single design point is removed when training GP emulator, i.e. the remaining 49 points are used for training. The trained GP emulator is then employed as a surrogate witin the MCMC random walk to obtain the posterior parameter distribution shown. Notice that the posterior distribution are often peaked near the truth values for $\alpha^{\rm eff}_s$ and $Q_s$. This procedure is repeated $8$ additional times, each time another random training point is removed, thus producing the remaining 8 panels in Fig.~\ref{fig:closure_exp_fluc_14}. In that figure, the RBF kernel augmented with a white noise kernel was used to produce results, for reasons that are explored in Sec.~\ref{sec:diff_kernels}.\footnote{Note however, that it is difficult to compare the posterior distributions with different shapes generated using different kernels and $N_{\rm PC}$.} Repeating this procedure for each parameter combination of the simulated model, as was done herein, ensures that the emulator doesn’t exhibit undesired behavior within the spanned parameter space.

\subsection{Selection of optimal emulator settings for Bayesian inference}
\label{sec:quantitative_closure_test}

In the previous section, while the closure tests at $9$ random design points seem to perform well, it is difficult quantify emulator performance by merely looking at posterior distributions, let alone compare the performance between different kernel functions and $N_{\rm PC}$. A performance measure of the GPE is thus devised leveraging the notion of scoring rules \cite{Matheson:1976,Winkler:1994,Gneiting:2007,Dawid:2014} in Bayesian statistics. Scoring rules are traditionally used for evaluating the accuracy of probabilistic predictive models \citep{Gneiting:2007}. In this study such a rule is used for a novel purpose of fitting emulator parameters for Bayesian analysis. The proposed measure, derived below, quantifies the amount of information loss induced by GPE modelling via an information-theoretic approach.

The main contribution to GPE information loss stems from the second moment of the posterior distribution (as will be shown below). We define a quantity $\Delta_d$ which measures the (second moment) deviation of the posterior distribution away from the truth value $x^{(d)}_{\rm truth}$ of the parameter $x$, while using the $d$-th design point as the truth:
\begin{equation}
    \Delta_d= \int \left(\frac{x-x^{(d)}_{\rm truth}}{x_{\rm max}-x_{\rm min}}\right)^2 p_d(x)dx,
\label{eq:Delta_d}
\end{equation}
where $|x_{\rm max}-x_{\rm min}|$ is the allowed range of a parameter to be constrained, and $p_d(x)$ is the marginalized posterior distribution obtained from the case where $d$-th design point is excluded. This is a new quantitative measure of the emulator's performance at recovering the truth from the mock data.

Using $\Delta_d$ defined as a closure test for one design point, averaging over all design points allows to obtain an overall performance of the emulator:
\begin{equation}
    \langle\Delta\rangle=\frac{1}{N_{\rm d}}\sum^{N_d}_{d=1} \Delta_d.
\end{equation}
In our study, the number of design points is $N_d=50$.

\begin{figure}
	\centering
	\includegraphics[width=0.5\textwidth]{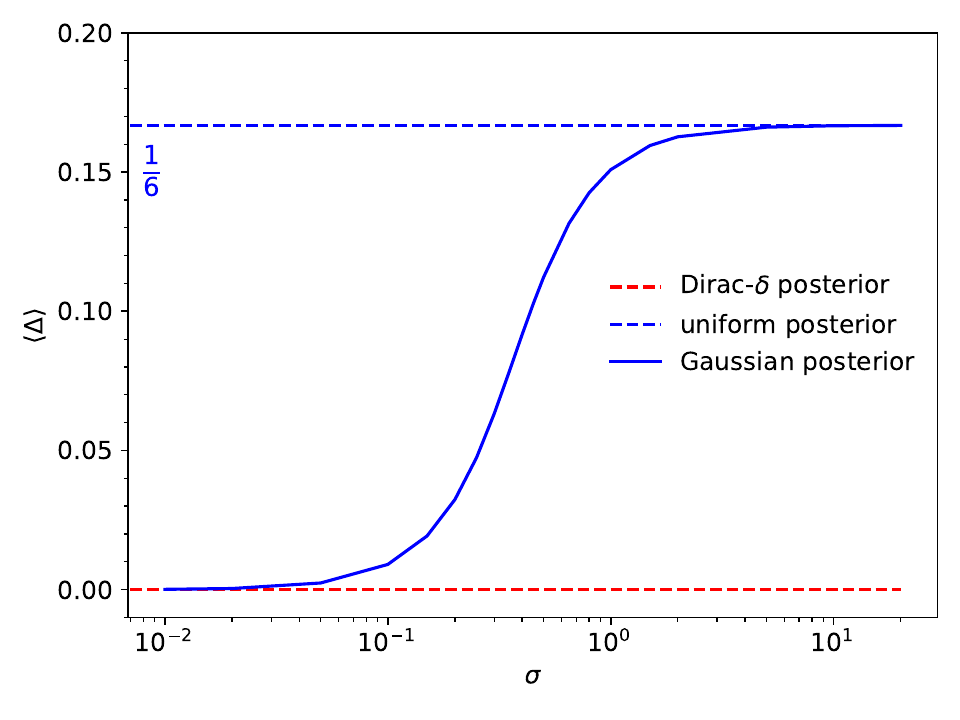}
        \vspace{-0.85cm}
	\caption{(Color online) $\langle\Delta\rangle$ calculated with Gaussian posterior centered at the truth versus the standard deviation $\sigma$ of the Gaussian posterior $p(x)$. The analytical result using a uniform posterior and Dirac-$\delta$ distribution posterior are shown with dashed lines.}
\label{fig:Delta_analytical}
\end{figure}

To understand how $\langle\Delta\rangle$ measures the deviation of the posterior distribution, the following example is illustrative. Suppose there are infinitely many design points uniformly distributed among the prior range. One calculates the values of $\langle\Delta\rangle$ with a uniform posterior $p(x)$ or a Gaussian posterior distribution centered at the truth with the variance being a free parameter to get a hint of the magnitude of $\langle\Delta\rangle$. For the case of a uniform $p(x)$ posterior, $\langle\Delta\rangle=1/6$ is immediately obtained. If one assumes a Gaussian posterior distribution centered at the truth, Fig.~\ref{fig:Delta_analytical} shows how $\langle\Delta\rangle$ changes as a function of $\sigma$: the standard deviation. $\langle\Delta\rangle$ approaches the value calculated with a uniform posterior when $\sigma\rightarrow\infty$, and goes to $0$ as $\sigma$ decreases. Thus, in closure tests, it is desirable for $\langle\Delta\rangle$ to be as close to 0 as possible\footnote{Note that when $\langle\Delta\rangle=0$, the GP emulator is indistinguishable from the full model.} while extracting the parameter values in the posterior, as that brings the sensitivity of the GP emulator to model-parameters closer to the sensitivity of the model itself to those parameters. 

We also recommend that, in practice, the averaging over $\Delta_d$ (yielding $\langle\Delta\rangle$) should be done using closure tests for cross-validation \citep{hastie2009elements}, rather than separating the simulation data set into a training and testing set for the GP emulator. The first approach via cross-validated closure tests ensures there is less variability in $\langle\Delta\rangle$ (and thus more stable fits for emulator parameters), since each data point is used for both training and testing (in different folds). The latter approach, which performs a single training-testing split of the simulated data, introduces greater variability in $\langle\Delta\rangle$ (especially for the 50 design points used herein), which in turn induces greater instability in model fitting. The latter can be of course be overcome by increasing the number of design points, but this is prohibitively expensive for our study.
%
\subsubsection{Connection to the Kullback-Leibler divergence}
\label{sec:connection_DKL}
The Kullback-Leibler Divergence \citep{cover1999elements} is defined between two probability distributions, the posterior $p(x)$ and the prior $q(x)$, as 
\begin{equation}
    D_{KL}(p|q)=\int p(x)\ln\left(\frac{p(x)}{q(x)}\right)dx.
\end{equation}
This provides an information-theoretic measure of how one probability distribution $p$ differs from a second probability distribution $q$, when the latter is used as a reference. In particular, for Bayesian analysis, this can be interpreted as a measure of information change when updating the prior distribution to a posterior distribution \citep{oladyshkin2019connection}. We restrict ourselves to a one dimensional system to ensure a clarity of our argument, with the generalization to multiple dimensions being straightforward. 

$D_{KL}$ is ill-defined (i.e., diverges) when the prior distribution $q(x)$ is a Dirac-$\delta$ function centered at $x_0$, the so-called ``true'' value. This can be seen using the limit representation $\delta(x-x_0)=\lim_{\sigma_q\rightarrow 0^+} \left[\frac{1}{\sqrt{\pi}\sigma_q}e^{-\left(x-x_0\right)^2/\sigma^2_q}\right]$ as: 
\begin{equation}\label{eqn:DKL_equivalence}
\begin{split}
    D_{KL}(p|q)&=\int dx\, p(x)\lim_{\sigma_q\rightarrow 0^+} \left[\ln\left(\frac{p(x)}{\frac{1}{\sqrt{\pi}\sigma_q}e^{-\left(x-x_0\right)^2/\sigma^2_q}}\right)\right] \\
    &= \lim_{\sigma_q\rightarrow 0^+}\left[\sigma^{-2}_q \int  p(x)(x-x_0)^2dx + \mathcal{O}\left(\ln \ \sigma_q \right) \right].\\
\end{split}
\end{equation}
However, noticing that the divergence is caused by the prior $q(x)$, with the posterior $p(x)$ being $\sigma_q$-independent, the non-singular part in Eq.~(\ref{eqn:DKL_equivalence}) is:
\begin{equation}
    \hat{\Delta}(x_0)=\int  p(x)(x-x_0)^2dx,
\label{eq:D_KL_2nd_moment}
\end{equation}
which bares resemblance to $\Delta_d$. Indeed, the shape of $\Delta_d(x_{\rm truth})$ in Eq.~(\ref{eq:Delta_d}) is the same as $\hat{\Delta}(x_0)$ in Eq.~(\ref{eq:D_KL_2nd_moment}), as required, the only difference being an $x$-independent overall normalization.  

The above derivation indicates that $\Delta_d$ is a measure of the finite part in $D_{KL}$ when the prior is a Dirac-$\delta$ distribution represented as the limit of a Gaussian distribution. $D_{KL}$ was not used before as a metric in closure test because of this divergence, which we have regulated. This connection is another motivation for using $\Delta_d$ and $\langle\Delta\rangle$ in closure tests. For a comparison with other metrics, see Ref.~\cite{Fan:2022ton}.

Notice that the interpretation of the Kullback-Leibler Divergence is different in closure tests compared to model-to-data Bayesian comparisons. In closure tests, the starting point is one where the information is maximized, i.e. the exact parameters are known {\it a priori}. However, since that parameters set is not used in the GP emulator, there is information loss which is captured by $D_{KL}$. In this case, a small $D_{KL}$ is desirable. 

Conversely, when using the trained GP emulators in Bayesian model-to-data comparisons, we are using a uniform prior distribution $q({\bf x})$ to extract the posterior distribution via Bayesian inference. In that case, the posterior has gained information relative to the prior and a larger $D_{KL}$ means parameters are better constrained.
\subsubsection{Comparison between different kernels}
\label{sec:diff_kernels}
\begin{figure*}[htb!]
	\centering
	\includegraphics[width=0.495\textwidth]{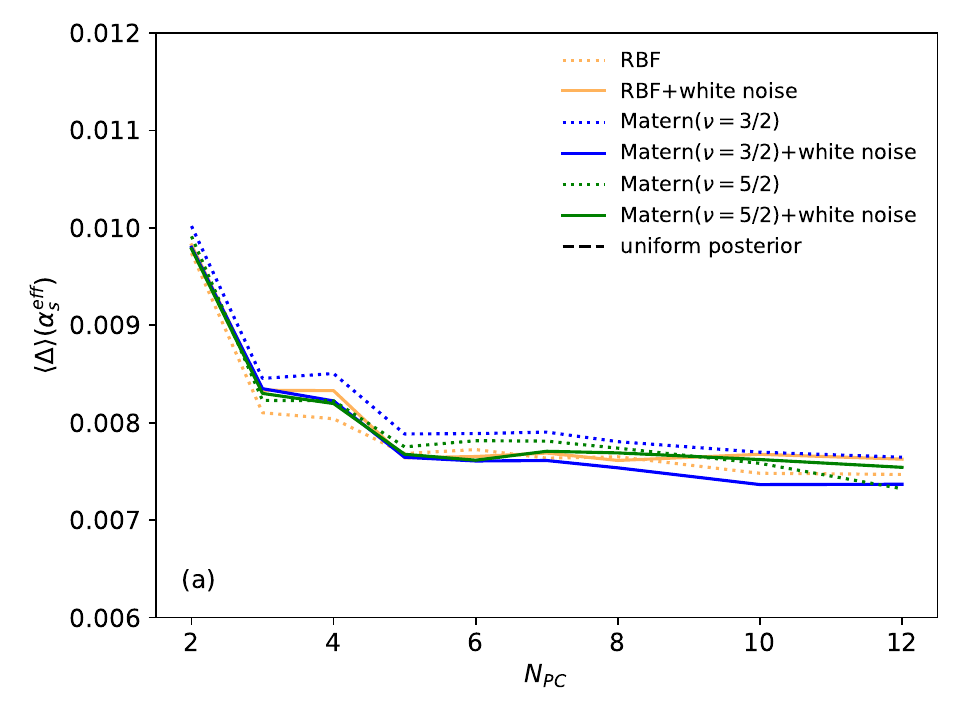}
	\includegraphics[width=0.495\textwidth]{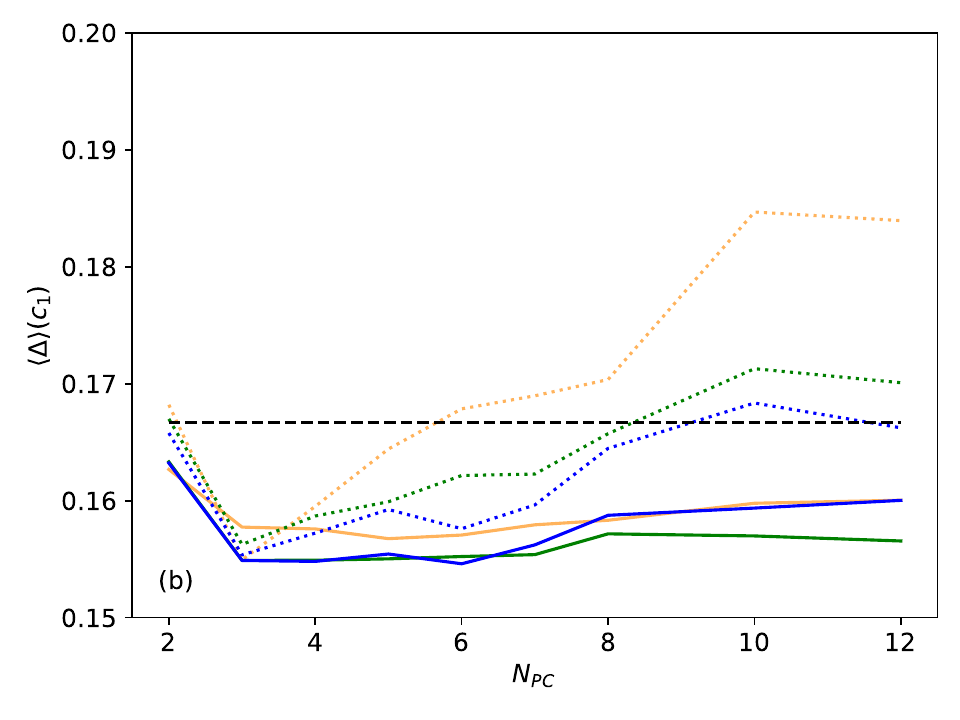}
	\includegraphics[width=0.495\textwidth]{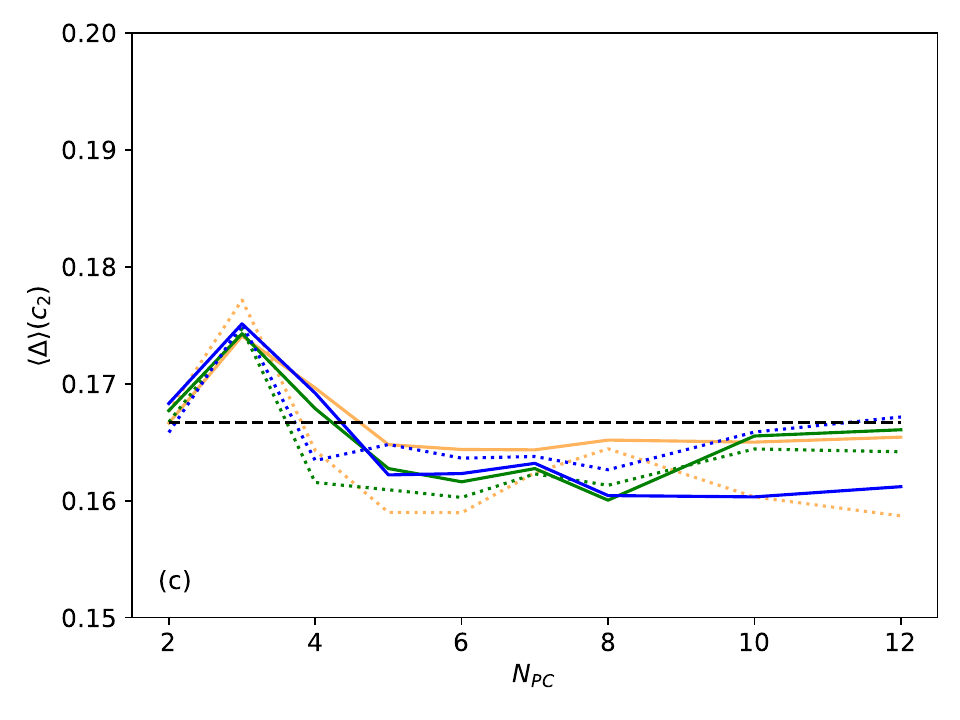}
	\includegraphics[width=0.495\textwidth]{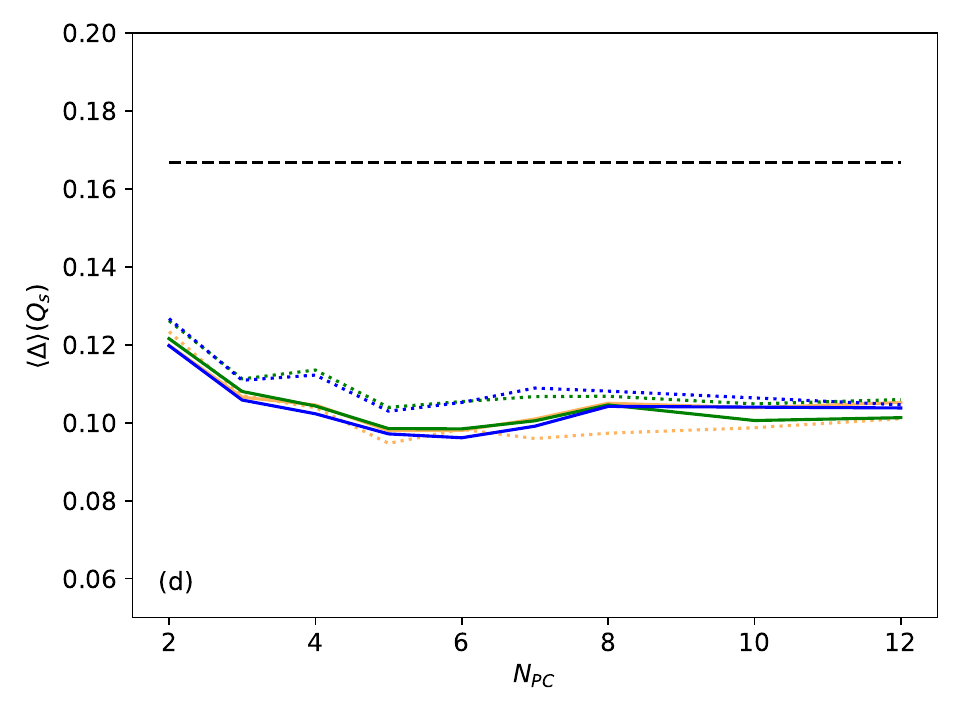}
        \vspace{-0.75cm}
	\caption{(Color online) Comparison of $\langle\Delta\rangle$ using different kernels and various number of principal components. The panels explore the sensitivity of $\langle\Delta\rangle$ to: $\alpha^{\rm eff}_s$ in (a), $c_1$ in (b), $c_2$ in (c), and $Q_s$ in (d).}
\label{fig:Delta_comparison} 
\end{figure*}

\begin{figure*}[htb!]
	\centering
	\includegraphics[width=0.495\textwidth]{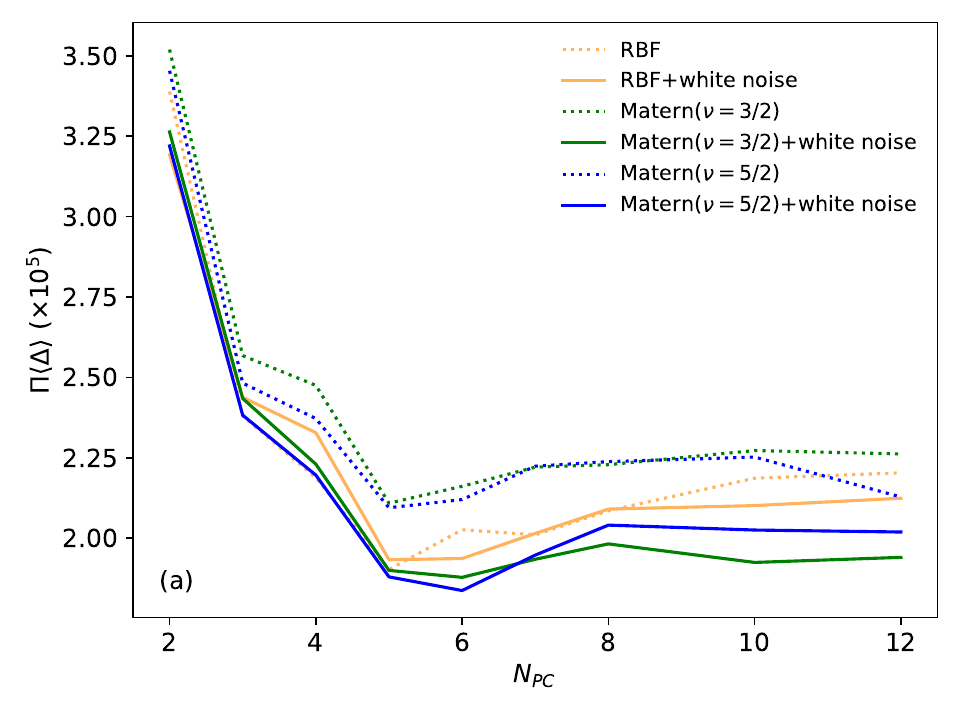}
	\includegraphics[width=0.495\textwidth]{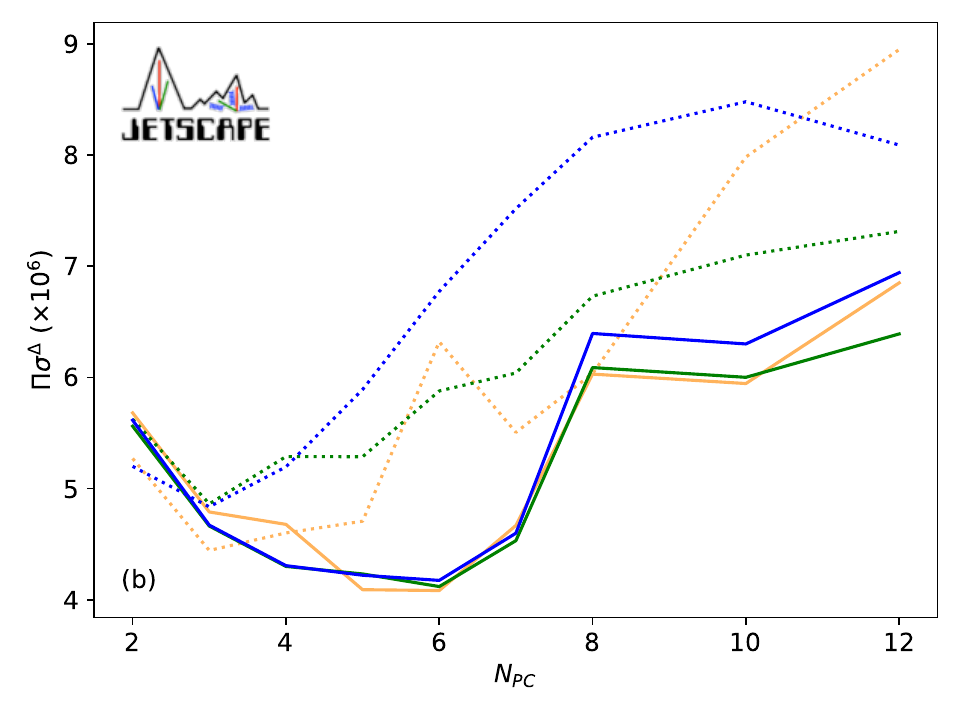}
        \vspace{-0.75cm}
	\caption{(Color online) Comparison $\Pi\langle\Delta\rangle$ (a) and $\Pi\langle\Delta\rangle$ (b) for different parameters with different kernels, training data selection, and number of principal components.}
\label{fig:Delta_comparison_all}
\end{figure*}

Six types of kernels are compared in this section: the RBF kernel, the Mat\'{e}rn ($\nu=5/2$), the Mat\'{e}rn ($\nu=3/2$), and linear combinations of these three kernels with the white noise kernel. The results are shown in Fig.~\ref{fig:Delta_comparison}. From the various panels therein, one can see that $\langle\Delta\rangle(\alpha^{\rm eff}_s)$ has the smallest values, and thus model dependence on $\alpha^{\rm eff}_s$ is best captured by the emulator. The second best emulated parameter is $Q_s$ as can be seein by $\langle\Delta\rangle(Q_s)$ in Fig.~\ref{fig:Delta_comparison}~(d). Finally, panels (b) and (c) of Fig.~\ref{fig:Delta_comparison} show that the emulator struggles to capture model sensitivity to $c_1$ and $c_2$, as is seen in $\langle\Delta\rangle(c_1)$ and $\langle\Delta\rangle(c_2)$.\footnote{Note  $\Delta=1/6$ is indicated by the blue (grey) dashed line, which is the analytical result assuming a uniform posterior.} This means the emulators are having trouble recovering these two parameters given the current level of uncertainties.
In order to pick the optimal settings for the emulator, the product of all the $\langle\Delta\rangle$ is investigated. That is, 
\begin{eqnarray}
\Pi \langle\Delta\rangle=\langle\Delta\rangle\left(\alpha^{\rm eff}_s\right)\langle\Delta\rangle(Q_s)\langle\Delta\rangle(c_1)\langle\Delta\rangle(c_2),
\end{eqnarray}
which is computed for different emulators in Fig.~\ref{fig:Delta_comparison_all}. As can be seen therein, a minimum in $\Pi\langle\Delta\rangle$ curve exists and gives the optimal choice of $N_{\rm PC}$. Also note that the white noise kernel improves the overall performance, thus highlighting the importance of including the white noise kernel in the GP emulator.  

One can also look at the variance of $\Delta$, i.e., $\sigma^\Delta$, which is calculated as:
\begin{eqnarray}
    \sigma^\Delta= \frac{1}{N_d}\sum^{N_d=50}_{d=1} (\Delta_d - \langle\Delta\rangle)^2.
\end{eqnarray}
Notice that $\sigma^\Delta$ is not an absolute indication of the emulator's performance, as it is possible to still have large mean and small variance (or small mean and large variance) at the same time (for a comparison with other metrics, see Ref.~\cite{Fan:2022ton} Sec. 6.4.1). Nevertheless, a large $\sigma^\Delta$ indicates that the emulator is not performing well on at least some of the design points. The product of all the $\sigma^\Delta$, which is
\begin{eqnarray}
\Pi \sigma^\Delta= \sigma^\Delta\left(\alpha^{\rm eff}_s\right) \sigma^\Delta(Q_s) \sigma^\Delta(c_1) \sigma^\Delta(c_2),
\end{eqnarray}
is shown in panel (b) of Fig.~\ref{fig:Delta_comparison_all}. Interestingly, the smallest values are also achieved at around $5$ or $6$ principal components, and with the kernels that include white noise. The decision to use $N_{\rm PC}=5$ is based on the results presented in Fig.~\ref{fig:Delta_comparison_all}. In this work, we will use the (RBF + white noise) kernel with $5$ principal components for its overall performance and consistency with previous studies.

The method discussed in this section is quite general and can easily be applied to other Bayesian analysis projects utilizing the Gaussian Process emulator as a surrogate model. Key ingredients for training the emulator, including choice of the kernel and the number of principal components, can all be determined by computing $(\langle\Delta\rangle,\Pi\langle\Delta\rangle)$ as well as ($\sigma^\Delta,\Pi\sigma^\Delta)$. The level of constraint on each parameter, can also be reflected by $\langle\Delta\rangle$. Note, however, that model-related (i.e. theoretical) systematic uncertainties have not yet been taken into account. One should also keep in mind that this is a measure of the average performance over the entire parameter space being searched. Currently, the average is the best estimate we can get since the truth (optimal) parameters are not known {\it a priori}. 
%
\subsubsection{Observable sensitivity analysis}
\label{sec:sensitivity_delta}
%
\begin{figure*}
	\centering
	\includegraphics[width=0.495\textwidth]{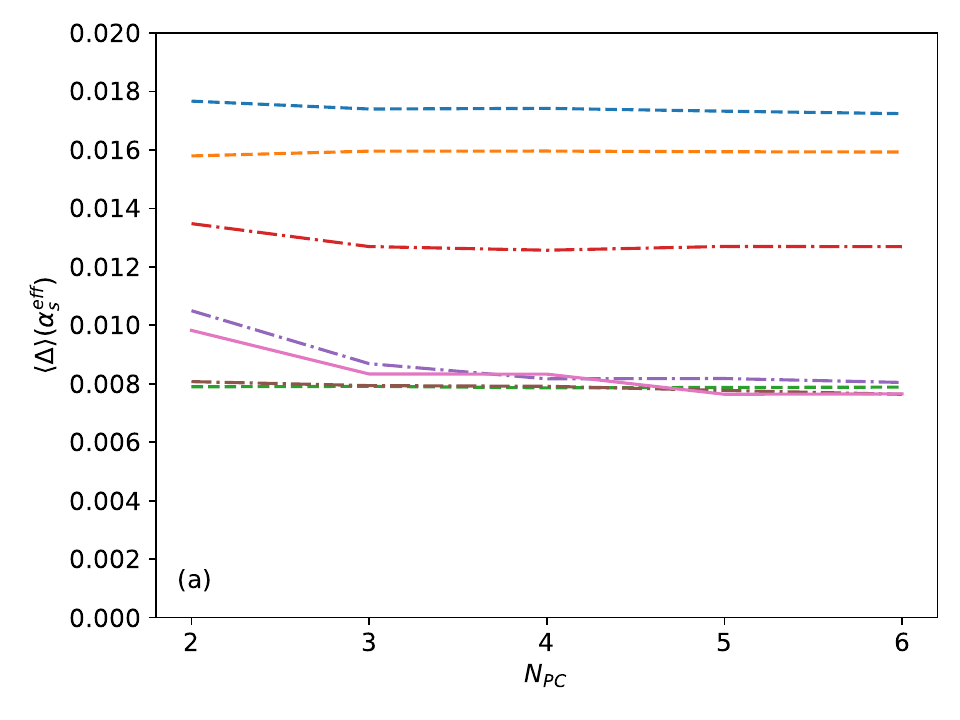}
	\includegraphics[width=0.495\textwidth]{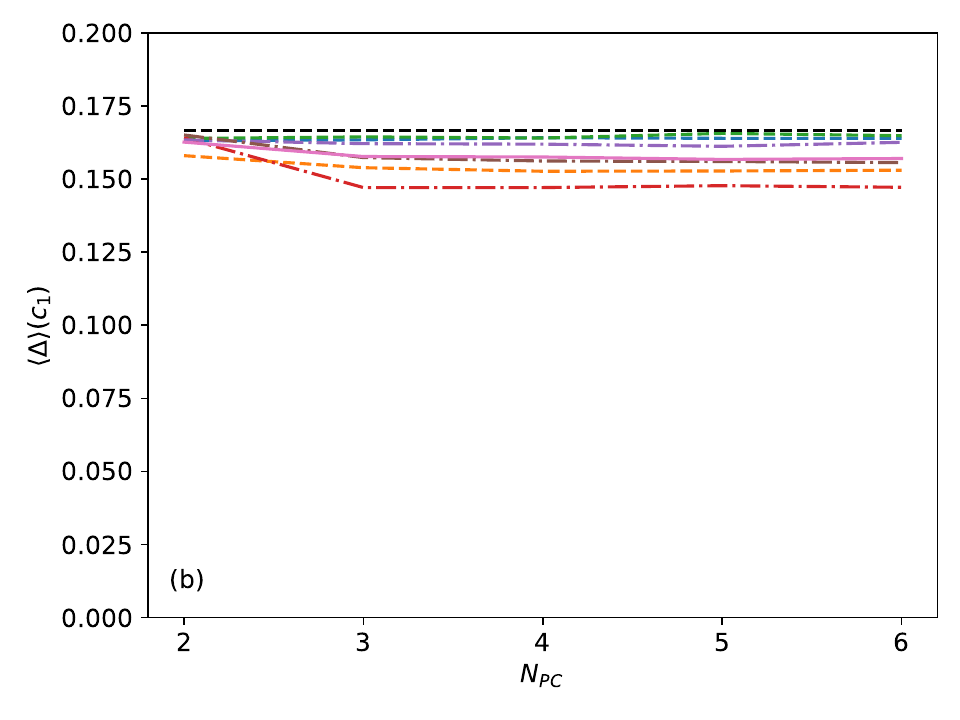}
	\includegraphics[width=0.495\textwidth]{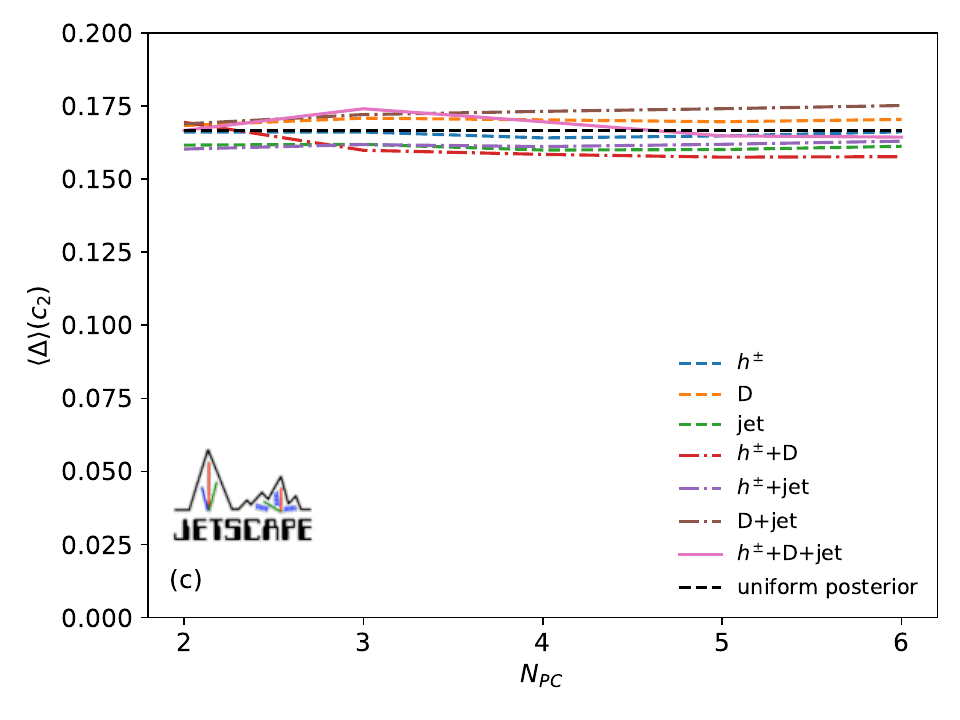}
	\includegraphics[width=0.49\textwidth]{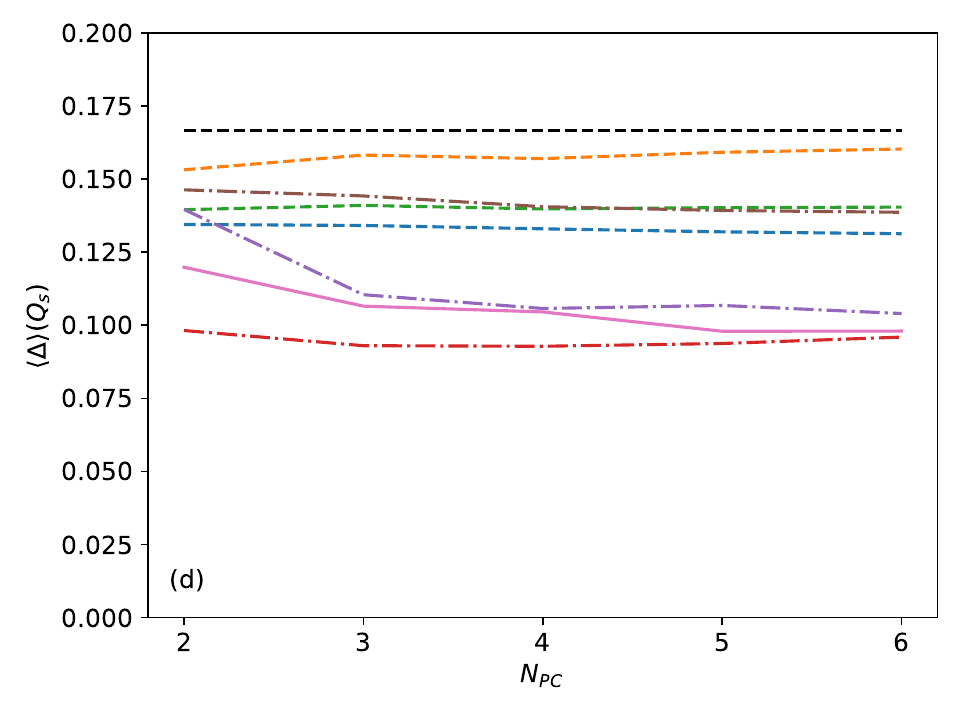}
        \vspace{-0.5cm}
    \caption{(Color online) The change of sensitivity of $\langle\Delta\rangle$ as different types of observables are considered. Panel (a) explores the change in sensitivity of $\langle\Delta\rangle\left(\alpha^{\rm eff}_s\right)$, (b) and (c) focus on $\langle\Delta\rangle(c_1)$, and $\langle\Delta\rangle(c_2)$, respectively, while (d) investigates the change in sensitivity for $\langle\Delta\rangle(Q_s)$ as various observables are incorporated.}
\label{fig:Delta_observable_sensitivity} 
\end{figure*}
$\langle\Delta\rangle$ can also be used to measure the sensitivity of model parameters to different observables. There are three categories of observables in our calibration: charged hadrons, D-mesons, and inclusive jet $R_{AA}$. The $\langle\Delta\rangle$ when calibrating to different combinations of these three types of observables are shown in Fig.~\ref{fig:Delta_observable_sensitivity}. We find that inclusive jet $R_{AA}$ observables are most sensitive to $\langle\Delta\rangle(\alpha^{\rm eff}_s)$, while charged hadron $R_{AA}$ governs the size of $\langle\Delta\rangle(Q_s)$. Of course, a combination of two or more of these observables improves $\langle\Delta\rangle$. No significant constraint can be derived on average for $c_1$ and $c_2$. The fact that $\Delta$ can be used to identify which observable(s) contribute the most to emulator prediction is a key feature guiding future emulator performance improvement. Together with the fact that $\Delta$ can also quantify the performance of GP emulation kernels as discussed in Sec.~\ref{sec:diff_kernels}, makes it an invaluable quantity to compute when training emulators.
%
\subsection{Bayesian Inference Results}
\label{sec:results}
%
\begin{figure*}
	\centering
	\includegraphics[width=\textwidth]{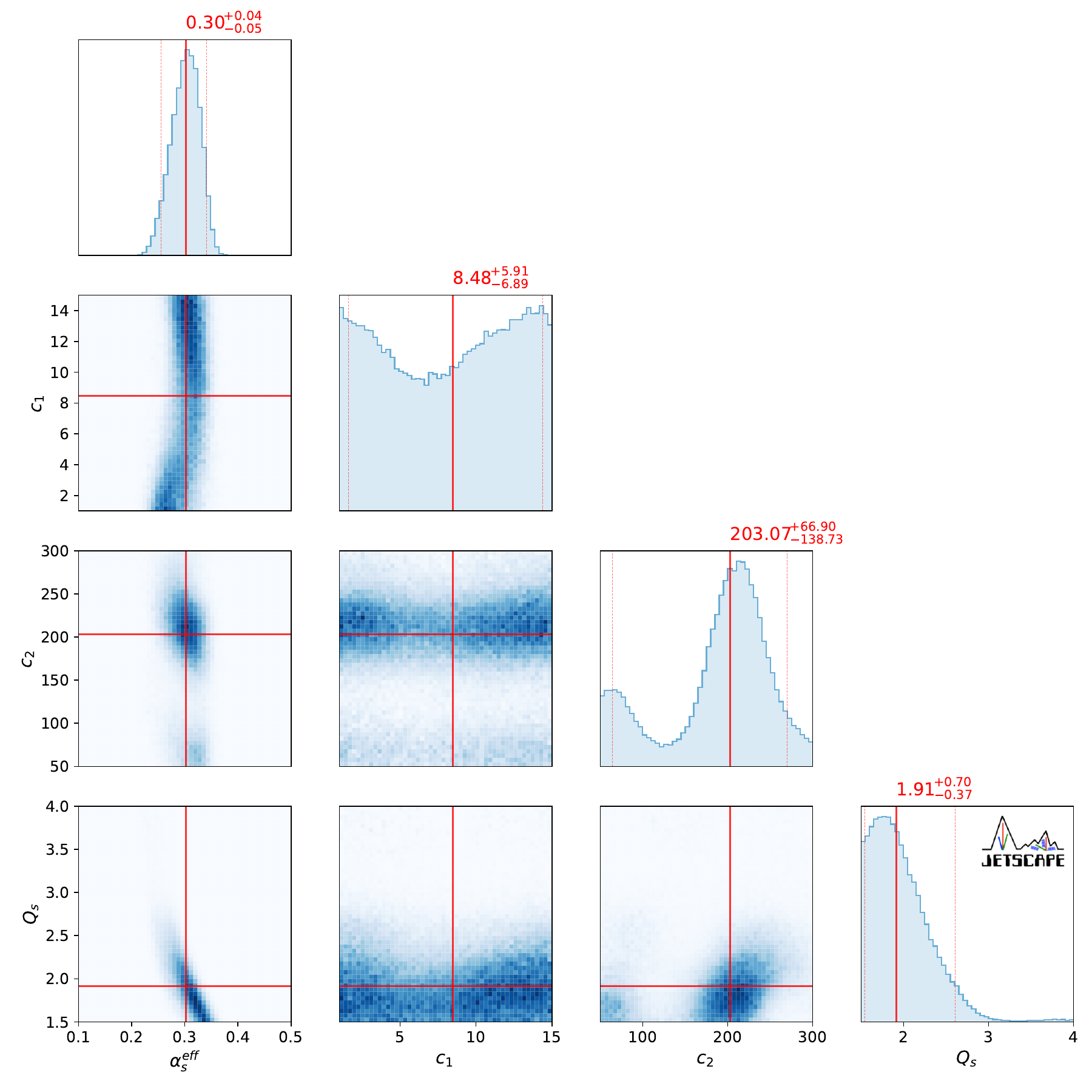}
        \vspace{-0.75cm}
	\caption{(Color online) The posterior distribution of model parameters. The emulator is using 5 PC and the RBF and white noise kernel combination and trained from 50 design points.}
\label{fig:BestPosteriorEstimation}  
\end{figure*}

\begin{figure*}
	\centering
	\includegraphics[width=\textwidth]{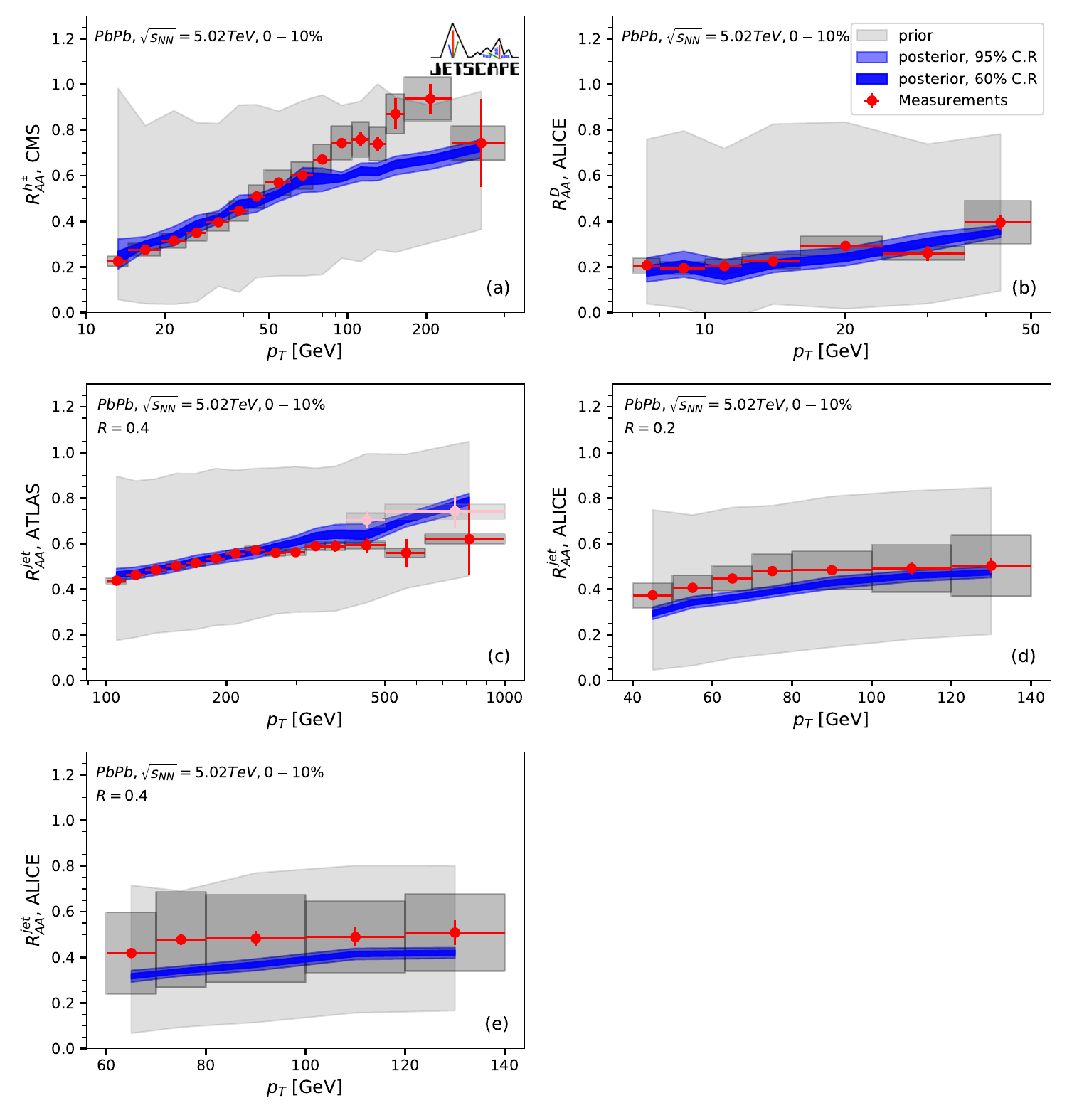}
        \vspace{-1cm}
	\caption{(Color online) Comparison between the posterior distribution of the observables and experiment data. Note that the pink (light grey) colored CMS data \cite{CMS:2021vui} in (c) are not used in training the GP emulator nor are they used in the subsequent Bayesian model-to-data comparison.}
\label{fig:BestObservablePosterior} 
\end{figure*}

\begin{figure}
	\centering
	\includegraphics[width=0.5\textwidth]{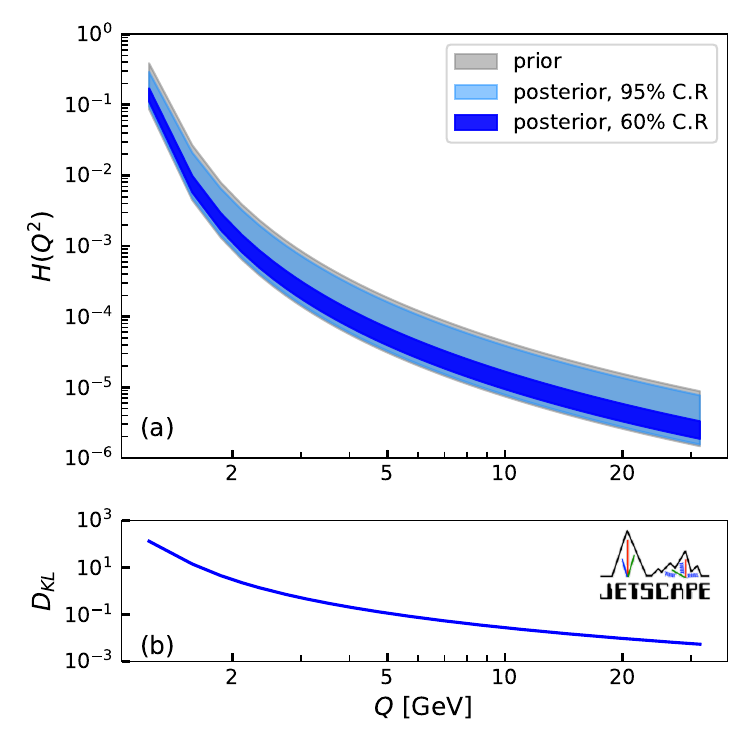}
        \vspace{-0.75cm}
	\caption{(Color online) Panel (a) depicts the prior (light grey), 95\% credible region of the posterior (light blue), and 60\% credible region of the posterior (deep blue) of $H(t)=H(Q^2)$ defined in Eq.~(\ref{eq:qhat_t}). Panel (b) displays the corresponding information gain using the Kullback-Leibler divergence $D_{KL}$).}
\label{fig:BestPosteriorHq}
\end{figure}
The posterior distribution of the parameters calibrating all five experimental data sets is shown in Fig.~\ref{fig:BestPosteriorEstimation}. The posterior distribution of the observables compared to data are shown in Fig.~\ref{fig:BestObservablePosterior}. Relative to the parameters used in Ref.~\cite{JETSCAPE:2022jer,JETSCAPE:2022hcb}, the posterior distribution in Fig.~\ref{fig:BestPosteriorEstimation} suggests similar value for $\alpha^{\rm eff}_s$ and a slightly smaller $Q_s$. Indeed, Ref.~\cite{JETSCAPE:2022jer,JETSCAPE:2022hcb} show that smaller values of $Q_s$ shifts the charged hadron and D-meson $R_{AA}$ upwards, a little closer to data compared to results found in Ref.~\cite{JETSCAPE:2022jer,JETSCAPE:2022hcb}, compensating somewhat for the mismatch in charged hadron $R_{AA}$ at high $p_T$ seen in our earlier study.

While the case of a virtuality independent $\hat{q}$ has been phenomenologically ruled out in Ref.~\cite{JETSCAPE:2022jer,JETSCAPE:2022hcb}, Fig.~\ref{fig:BestPosteriorEstimation} shows a weak constraint on $c_1$ and a slight preference for large $c_2$ values. An intuitive explanation for why $c_1$ and $c_2$ are hard to constrain is that the energy loss in the MATTER regime goes to zero when $c_1$ and $c_2$ go to infinity. Furthermore, MATTER's energy loss is quickly reduced when $c_1$ and $c_2$ values are non-negligible as is the case herein, and was shown in Ref.~\cite{JETSCAPE:2022jer,JETSCAPE:2022hcb} using much higher statistics.\footnote{$\langle\Delta\rangle(c_1)$ and $\langle\Delta\rangle(c_2)$ also reflect the small sensitivity of $(c_1,c_2)$ to observables.} In the future, a measurement with smaller uncertainties at high-$p_T$ can be beneficial in helping better constrain $c_1$ and $c_2$, together with higher statistics of theoretical calculations to improve D-meson $R_{AA}$ predictions. 

In Fig.~\ref{fig:BestPosteriorHq}, shows the constraint on $H(Q^2)$ resulting from this Bayesian analysis. Panel (a)  shows the prior, 95\% credible region, and 60\% credible region of $H(t)=H(Q^2)$ as a function of $Q$. The Kullback-Leibler divergence (defined in Section \ref{sec:connection_DKL}) depicted in Fig.~\ref{fig:BestPosteriorHq} (b) monotonically decreases with $Q$. This is likely due to the fact that the second-order term and the fourth-order term in $H(Q^2)$ are comparable in magnitude in this region. Thus, the {\it joint distribution} of $c_1$ and $c_2$ at low $Q$ is constrained, rather than their individual distributions.  

\subsection{Sensitivity to different observables}

\begin{figure*}
	\centering
	\includegraphics[width=0.325\textwidth]{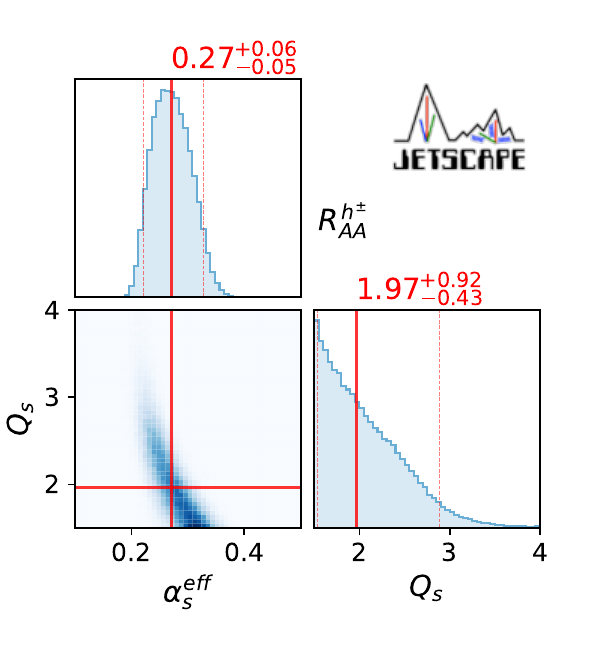}
	\includegraphics[width=0.325\textwidth]{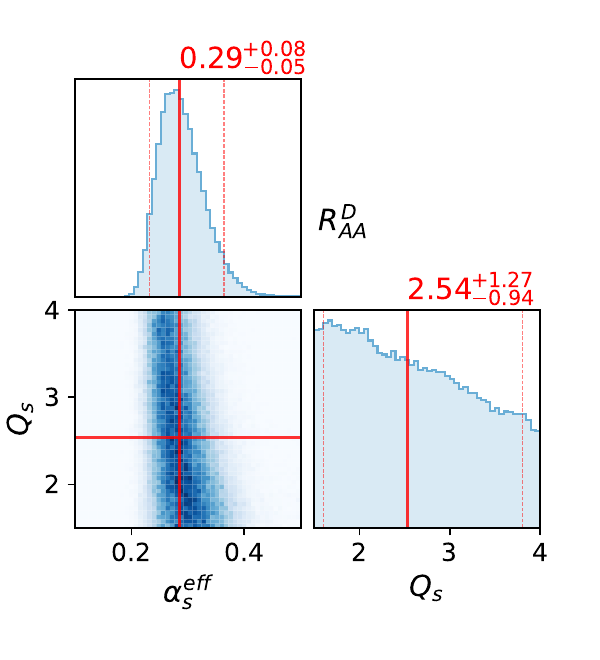}
	\includegraphics[width=0.325\textwidth]{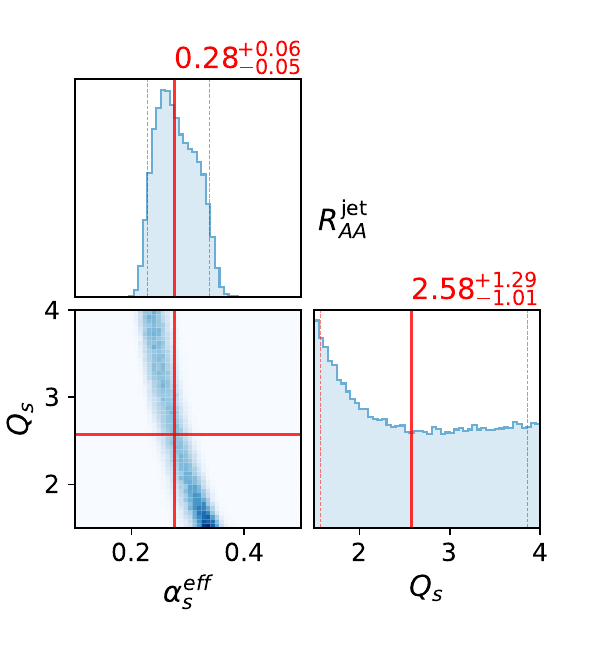}
	\includegraphics[width=0.325\textwidth]{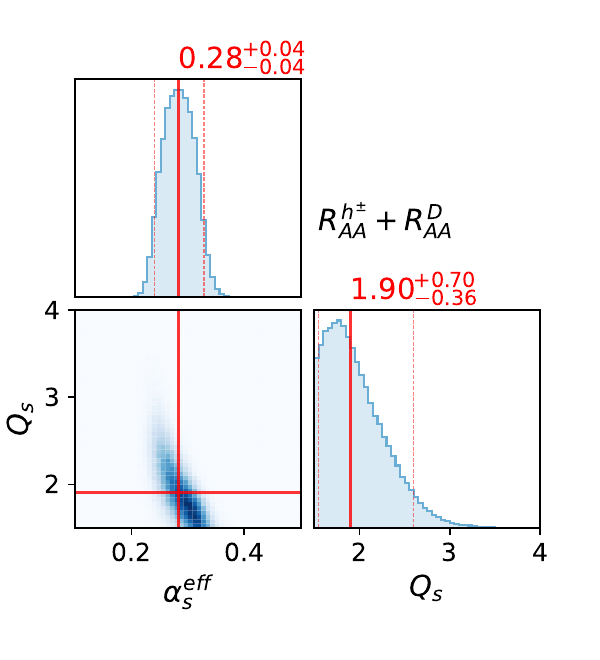}
	\includegraphics[width=0.325\textwidth]{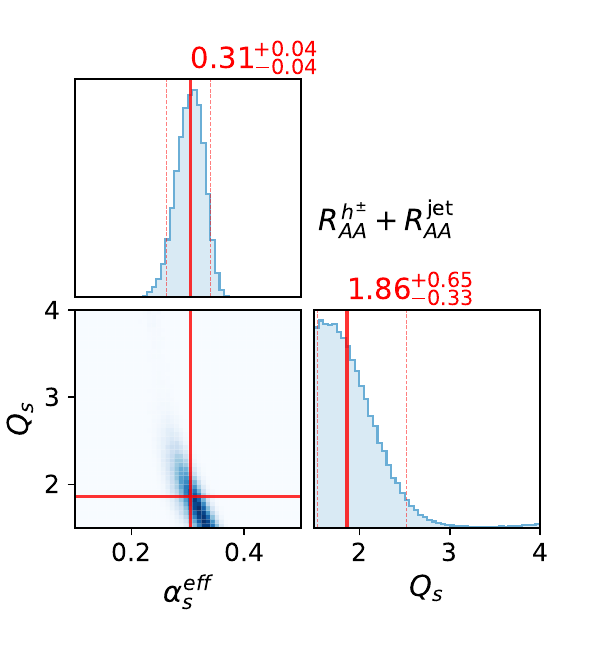}
	\includegraphics[width=0.325\textwidth]{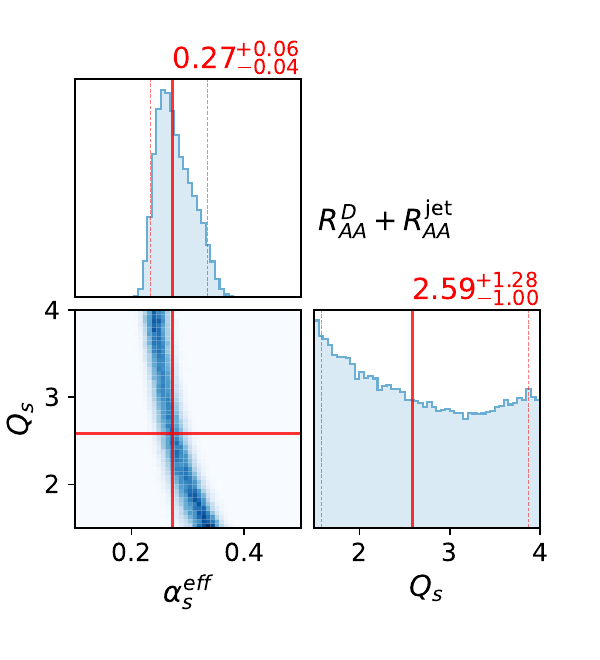}
        \vspace{-0.5cm}
	\caption{(Color online) Posterior distribution of $\alpha^{\rm eff}_s$ and $Q_s$, calibrated using different observables individually as well as pairwise.}
\label{fig:posterior_sensitivity}
\end{figure*}

We analyse in Fig.~\ref{fig:posterior_sensitivity} how different observables (and observable combinations), affect the constraints presented in Fig.~\ref{fig:BestPosteriorEstimation}. Since the constraint for $c_1$ and $c_2$ are weak, only the posterior distribution of $\alpha^{\rm eff}_s$ and $Q_s$ are shown in Fig.~\ref{fig:posterior_sensitivity}. One can see that charged hadron data are the main reason why large $Q_s$ is disfavored. Indeed, looking at the bottom panels in Fig.~\ref{fig:posterior_sensitivity} one sees that any combination including charged hadron $R_{AA}$ provides a stronger sensitivity to $Q_s$, compared to using jet and D-mesons, thus furthering our undersanding of the results in Fig.~\ref{fig:BestPosteriorEstimation}. Looking at the off-diagonal joint distribution plots in Fig.~\ref{fig:posterior_sensitivity} also reveals that $\alpha^{\rm eff}_s$ and $Q_s$ are anti-correlated; an observation that may be useful in future Bayesian analysis.
%
\section{Conclusion}
\label{sec:conclusion}
In this work, we performed Bayesian inference for a multistage parton energy loss approach in heavy-ion collisions. The model calculation is calibrated to charged hadron, D-meson, and inclusive jet $R_{AA}$ measurements. The challenge, however, is the sizable uncertainties in both experimental measurements and model calculation (emulation). We have validated our Bayesian workflow in both the forward direction, i.e., mapping from model parameters to observables, and the inverse direction, in the context of closure tests. Specifically, we have proposed the $\langle\Delta\rangle$ metric in cross-validation closure tests which we connected to the Kullback-Leibler divergence. With this metric, the optimal settings for emulating the model calculation was determined. The $\langle\Delta\rangle$ metric can also give guidance about the sensitivity of each parameter to different observables. Its usefulness for future Bayesian inference studies is what makes $\langle\Delta\rangle$ an important quantity to calculate, and is the main result of our work. 

The results of our Bayesian analysis finds optimal values for $\alpha^{\rm eff}_s$ and $Q_s$ to be similar to what have been used in previous studies \cite{JETSCAPE:2022jer,JETSCAPE:2022hcb,Fan:2022ton}, though now these values are on a much firmer footing. The constraints for $c_1$ and $c_2$ are much weaker due to their small sensitivity to the current observables. The small sensitivity is verified by both the $\langle\Delta\rangle$ metric and arguments about the influence of $c_1$ and $c_2$ on parton energy loss \cite{Fan:2022ton}. Another outcome of our Bayesian analysis is the identification that $\alpha^{\rm eff}_s$ is most sensitive to inclusive jet $R_{AA}$, while $Q_s$ is most sensitive to charged hadron $R_{AA}$. 

While there are sizeable experiment uncertainties in the data against which our model was calibrated, there are also sizeable theoretical uncertainties owing, in part, to the computation resources available for simulation. The latter restricted us to a subset of experimental data included in our Bayesian study. To improve theoretical uncertainties, one could adopt a more complex parametrization for $\hat{q}$ to account for the potential difference between data and theory at high $p_T$. One could also sample $c_1$ and $c_2$ on a logarithmic scale to improve the design point density at large $c_1$ and $c_2$. Our calibration should also include more observables, different collision energies and systems, as well as centralities. The JETSCAPE Collaboration has an ongoing Bayesian analysis that will address most of these points \cite{Ehlers:2022ulm}. Furthermore, we expect that improved experimental uncertainties, such as releasing the full covariance matrix, would benefit future Bayesian studies. Thus, a more detailed discussion of the uncertainties for both experiment (e.g. off-diagonal experimental systematic uncertainty) and theory (systematic model uncertainty), should be included in a future study. 

\section*{Acknowledgments}
\label{Ack}

This work was supported in part by the National Science Foundation (NSF) within the framework of the JETSCAPE collaboration, under grant number OAC-2004571 (CSSI:X-SCAPE). It was also supported under ACI-1550172 (Y.C. and G.R.), ACI-1550221 (R.J.F. and M.K.), ACI-1550223 (U.H., L.D., and D.L.), ACI-1550225 (S.A.B., T.D., W.F.), ACI-1550228 (J.M., B.J., P.J., X.-N.W.), and ACI-1550300 (S.C., A.K., J.L., A.M., C. M., H.M., T. M., C.N., J.P., L.S., C.Si., I.S., R.A.S. and G.V.); by PHY-1516590 and PHY-1812431 (R.J.F., M.K., C. P. and A.S.), by PHY-2012922 (C.S.); it was supported in part by NSF CSSI grant number \rm{OAC-2004601} (BAND; D.L. and U.H.); it was supported in part by the US Department of Energy, Office of Science, Office of Nuclear Physics under grant numbers \rm{DE-AC02-05CH11231} (X.-N.W.), \rm{DE-AC52-07NA27344} (A.A., R.A.S.), \rm{DE-SC0013460} (S.C., A.K., A.M., C.S., I.S. and C.Si.), \rm{DE-SC0021969} (C.S. and W.Z.), \rm{DE-SC0004286} (L.D., U.H. and D.L.), \rm{DE-SC0012704} (B.S.), \rm{DE-FG02-92ER40713} (J.P.) and \rm{DE-FG02-05ER41367} (T.D., W.F., J.-F.P., D.S. and S.A.B.). The work was also supported in part by the National Science Foundation of China (NSFC) under grant numbers 11935007, 11861131009 and 11890714 (Y.H. and X.-N.W.), under grant numbers 12175122 and 2021-867 (S.C.), by the Natural Sciences and Engineering Research Council of Canada (C.G., M.H., S.J., and G.V.),  by the University of Regina President's Tri-Agency Grant Support Program (G.V.), by the Office of the Vice President for Research (OVPR) at Wayne State University (Y.T.), and by the S\~{a}o Paulo Research Foundation (FAPESP) under projects 2016/24029-6, 2017/05685-2 and 2018/24720-6 (A. L. and  M.L.). U.H. would like to acknowledge support by the Alexander von Humboldt Foundation through a Humboldt Research Award. C.S. acknowledges a DOE Office of Science Early Career Award. 
Computations were carried out on the National Energy Research Scientific Computing Center (NERSC), a U.S.Department of Energy Office of Science User Facility operated under Contract No. DE-AC02-05CH11231. The bulk medium simulations were done using resources provided by the Open Science Grid (OSG) \cite{Pordes:2007zzb, Sfiligoi:2009cct}, which is supported by the National Science Foundation award \#2030508. Data storage was provided in part by the OSIRIS project supported by the National Science Foundation under grant number OAC-1541335.

\appendix
\section{Stability of the posterior to fluctuations}
\label{appdx:1}
The closure test offers an excellent check that the emulator can reproduce mock data (model calculation) at many design points. However, the relation between the posterior distribution and the uncertainty level of the training data is still not explored. It is difficult to reduce the fluctuations of the training data as it requires running more events. The other direction is easier to explore and can be investigated in two ways.

\begin{figure*}
	\centering
	\includegraphics[width=\textwidth]{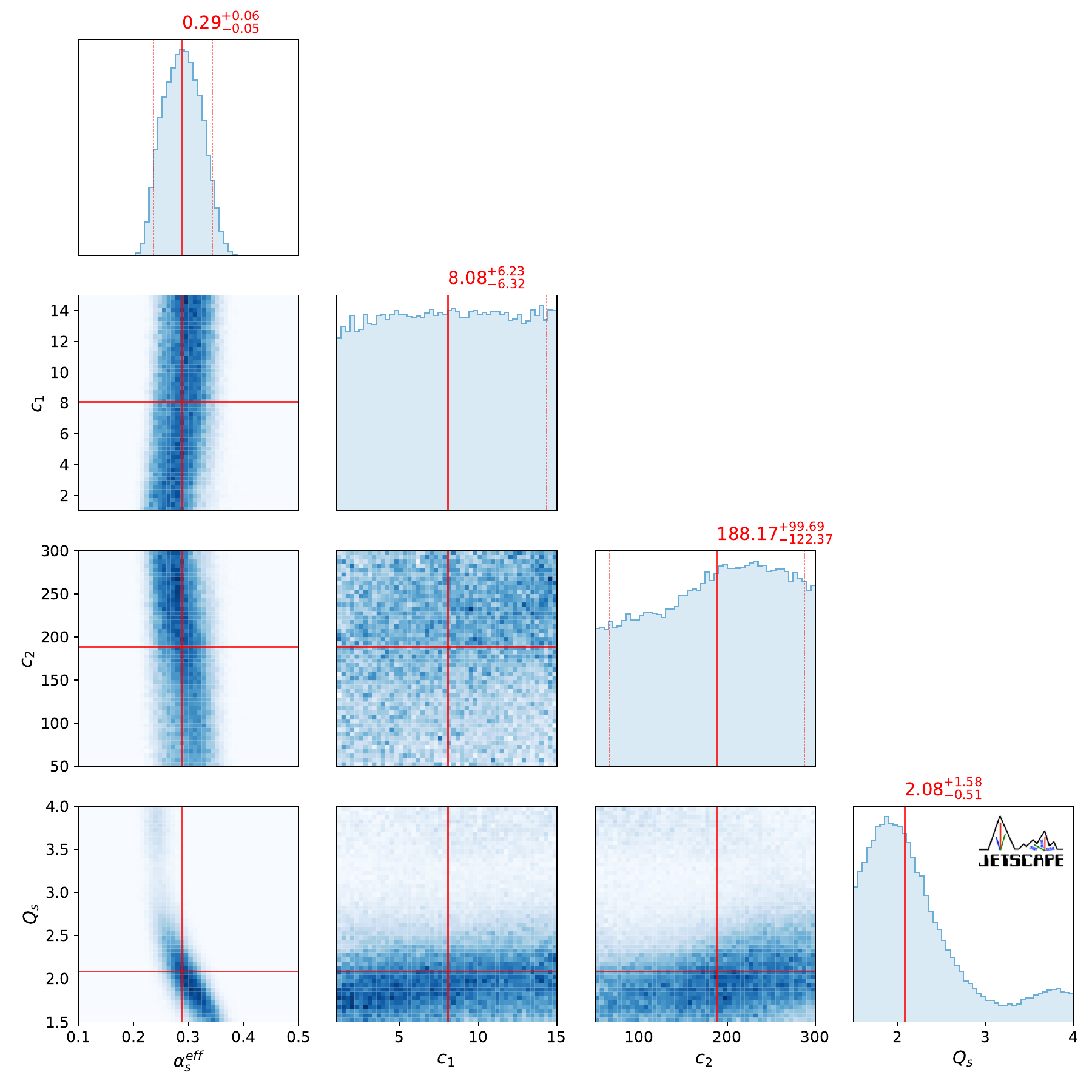}
        \vspace{-0.75cm}
	\caption{(Color online) The posterior distribution of the model parameters. Only $1/3$ of events are used for each design point.}
\label{fig:PosteriorEstimation_reduce_stat}
\end{figure*}

First, we can reduce the statistics when generating the training data, which we have done by only using $1/3$ of events at each design point. Thus, the effects of statistical fluctuations for each observable are scaled up and can be readily appreciated. Figure~\ref{fig:PosteriorEstimation_reduce_stat} shows the posterior distribution of the parameters. Compared to what is shown in Fig.~\ref{fig:BestPosteriorEstimation}, the posterior distribution is not wildly altered in Fig.~\ref{fig:PosteriorEstimation_reduce_stat}, except for weaker constraint on $c_1$ and $c_2$, which remain the roughly the same.
The effects of increasing the statistics by a factor of three significantly improve the constraint on the $\alpha^{\rm eff}_s$ and $Q_s$ parameters, while the effects on $(c_1,c_2)$ are more modest.

\begin{figure*}
	\centering
	\includegraphics[width=0.48\textwidth]{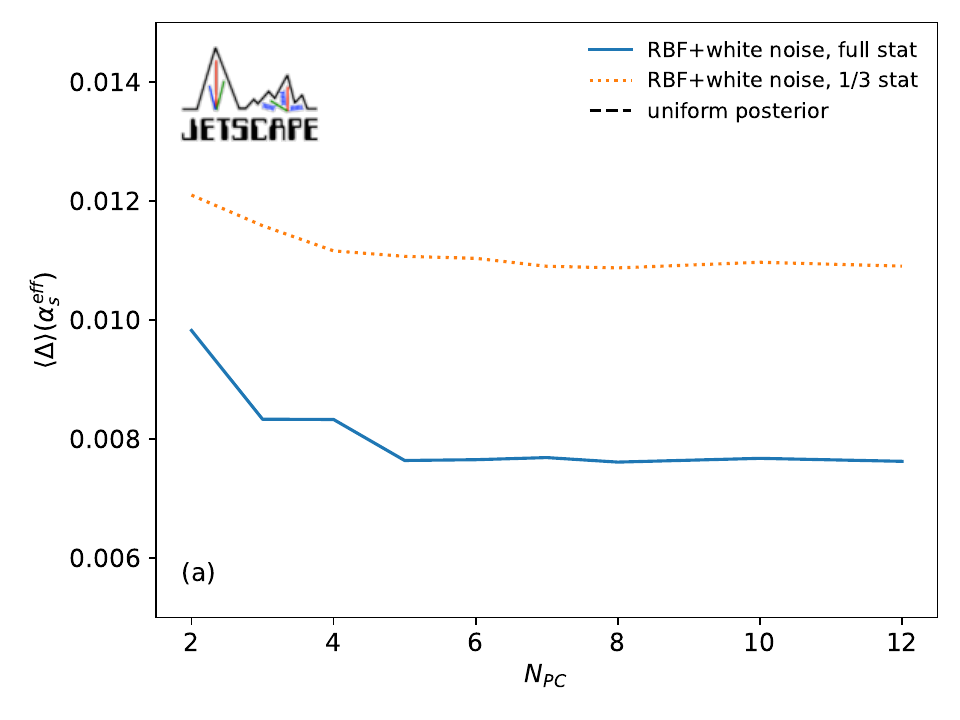}
	\includegraphics[width=0.48\textwidth]{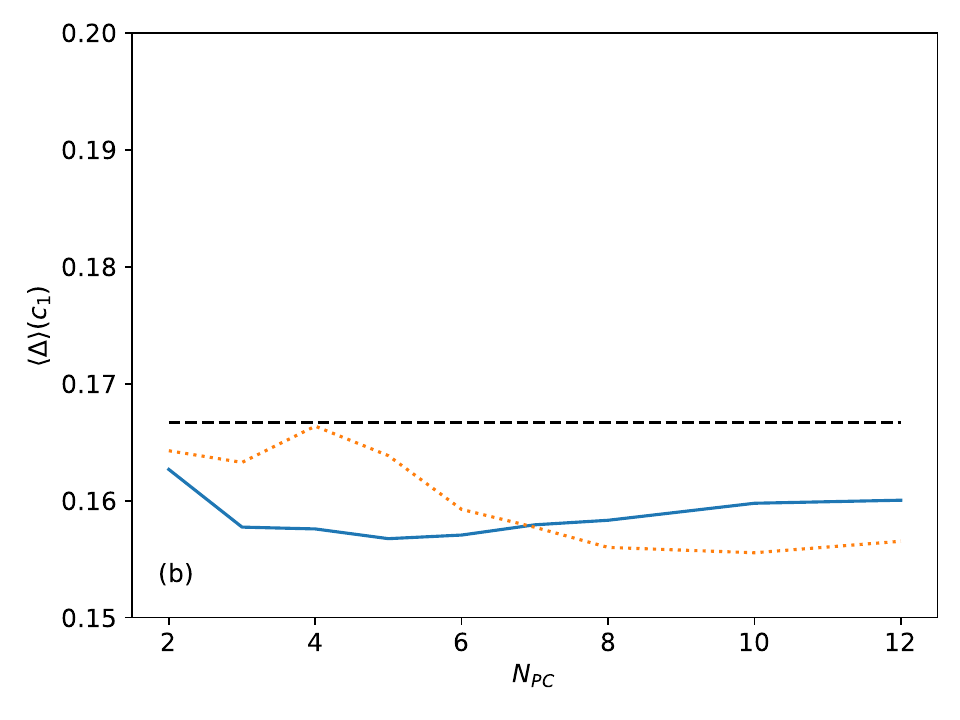}
	\includegraphics[width=0.48\textwidth]{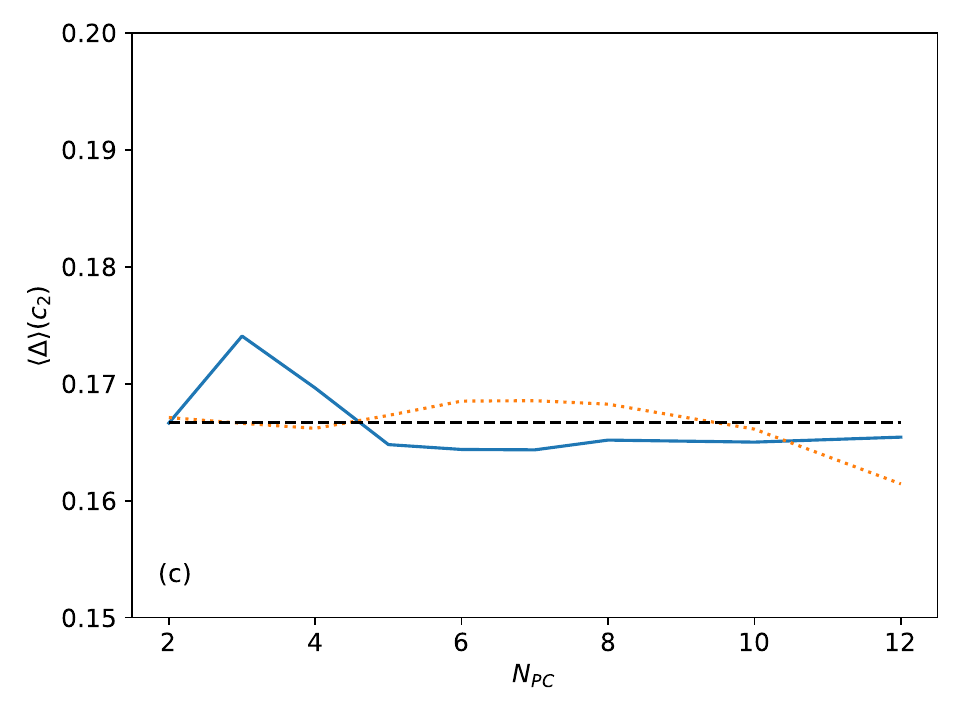}
	\includegraphics[width=0.48\textwidth]{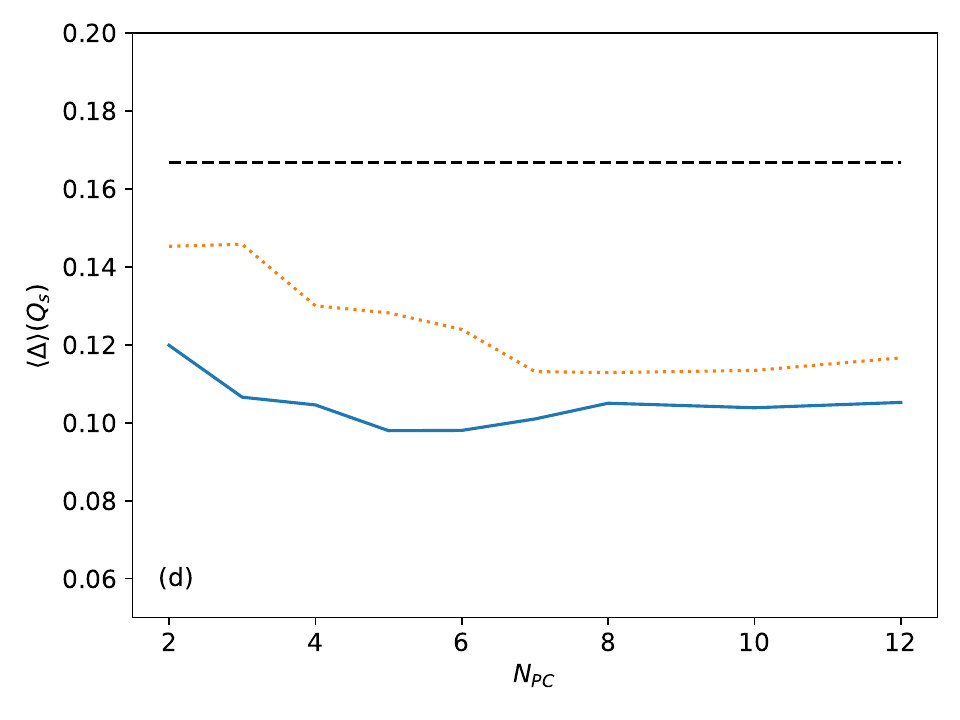}
        \vspace{-0.5cm}
	\caption{(Color online) Comparison of $\langle\Delta\rangle$ for different parameters with the same settings for the emulator but different statistics for the training data.}
\label{fig:Delta_comparison_stat} 
\end{figure*}

\begin{figure*}
	\centering
	\includegraphics[width=\textwidth]{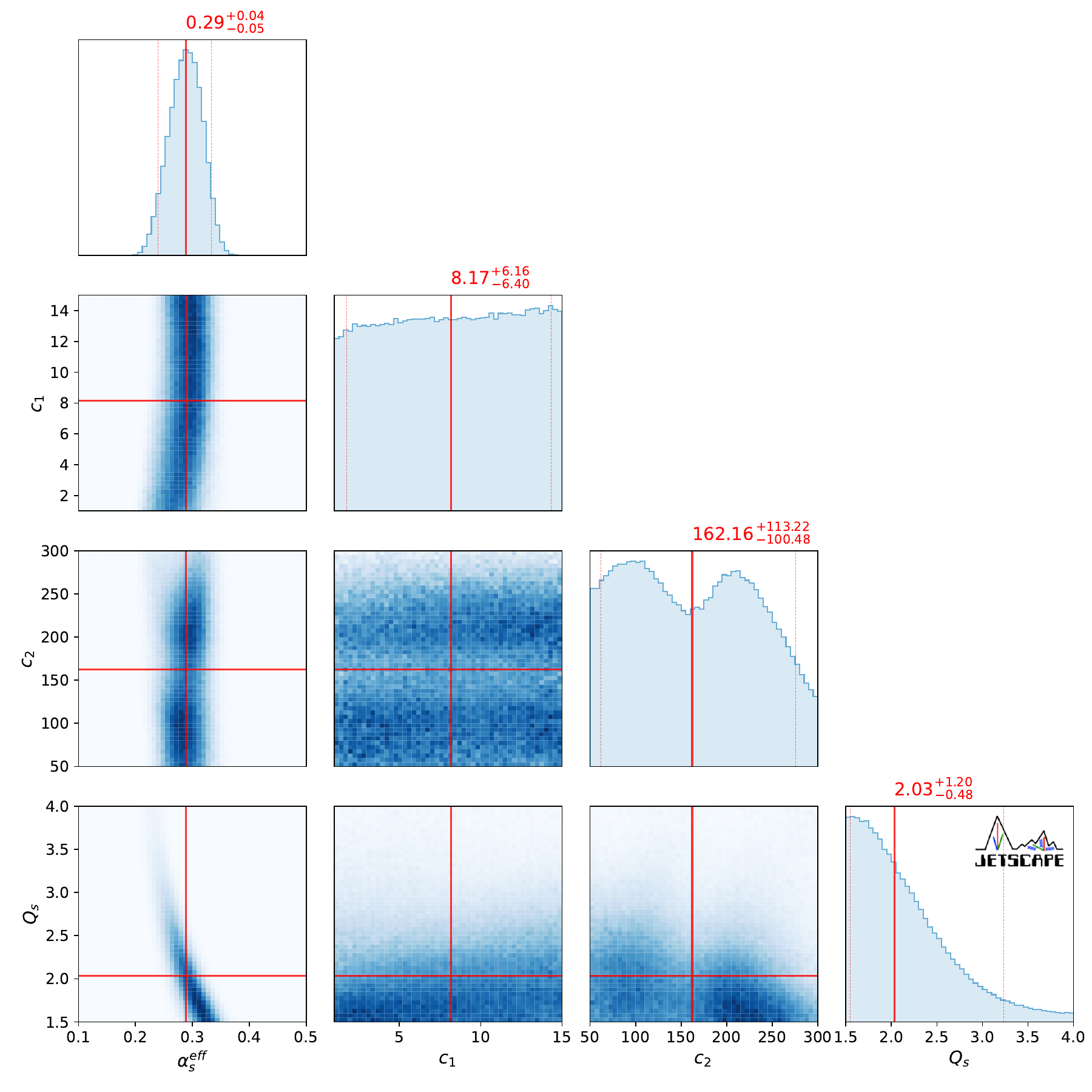}
        \vspace{-0.75cm}
	\caption{The posterior distribution of the model parameters with added Gaussian noise to all training data. The Gaussian noise has a mean $\mu=0$ and standard deviation $\sigma=0.05$.}
 \label{fig:PosteriorEstimation_more_fluc_0.05}
\end{figure*}

In Fig.~\ref{fig:Delta_comparison_stat}, we study how improved statics affect $\langle\Delta\rangle$. A marked improvement is seen on $\langle\Delta\rangle(\alpha^{\rm eff}_s)$ and $\langle\Delta\rangle(Q_s)$, while $\langle\Delta\rangle$ for $c_1$ and $c_2$ is modest, hovering around $1/6$, showing that constraints on $c_1$ and $c_2$ almost do not change when tripling the statistics. More statistics in D-meson $R_{AA}$ calculations, though a different parametrization for $\hat{q}(t)$ maybe needed and is being explored in Ref.~\cite{Ehlers:2022ulm}. Of course, reduced experimental uncertainties will also improve this situation.
Finally, the last element we considered is adding more Gaussian noise to all the model calculations for each design point. That is, every observable gets an equal amount of additional statistical fluctuation. In Fig.~\ref{fig:PosteriorEstimation_more_fluc_0.05}, we can see the updated posterior distribution when an additional noise with $0.05$ standard deviation is introduced. Similar posterior distributions of the parameters are seen compared to Fig.~\ref{fig:BestPosteriorEstimation} and Fig.~\ref{fig:PosteriorEstimation_reduce_stat}. 

\bibliography{references}
\end{document}